\documentclass[pre,aps,12pt]{revtex4}
\usepackage{graphicx}

\begin{document}

\title{Delta chain with anisotropic ferromagnetic and
antiferromagnetic interactions}
\author{D.~V.~Dmitriev}
\author{V.~Ya.~Krivnov}
\email{krivnov@deom.chph.ras.ru}

\affiliation{Institute of Biochemical Physics of RAS, Kosygin str.
4, 119334, Moscow, Russia.}
\date{}

\begin{abstract}
We consider analytically and numerically an anisotropic
spin-$\frac{1}{2}$ delta-chain (sawtooth chain) with
nearest-neighbor ferromagnetic and next-nearest-neighbor
antiferromagnetic interactions. For certain values of the
interactions a lowest one-particle band becomes flat and there is
a class of localized-magnon eigenstates which form a ground state
with a macroscopic degeneracy. In this case the model depends on a
single parameter which can be chosen as the anisotropy of the
exchange interactions. When this parameter changes from zero to
infinity the model interpolates between the one-dimensional
isotropic ferromagnet and the frustrated Ising model on the
delta-chain. It is shown that the low-temperature thermodynamic
properties in these limiting cases are governed by the specific
structure of the excitation spectrum. In particular, the specific
heat has one or infinite number of low-temperature maxima for the
small or the large anisotropy parameter, correspondingly.
Numerical calculations of finite chains demonstrate that this
behavior is generic for definite values of the anisotropy
parameter.
\end{abstract}

\maketitle

\section{Introduction}

\begin{figure}[tbp]
\includegraphics[width=3in,angle=0]{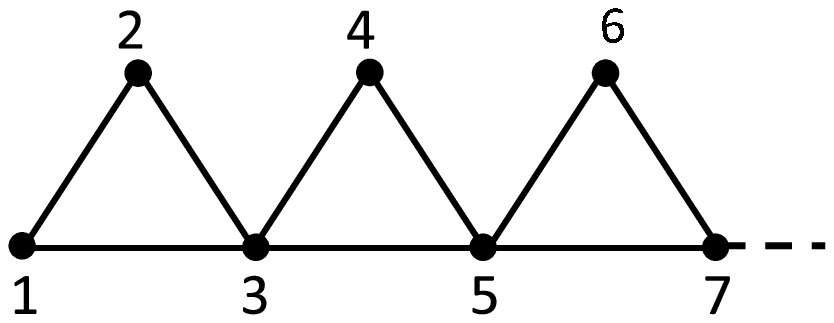}
\caption{The $\triangle$-chain model.}
\label{triangles}
\end{figure}

Quantum many-body systems with a single-particle flat-band have
attracted much attention
\cite{flat2008,Wu,Wang,Sun,Shulen,Zhit,Mak}. Frustrated quantum
spin systems represent examples where flat-band physics may lead
to new interesting phenomena such as a nonzero residual ground
state entropy, extra low-temperature peak in the specific heat etc
\cite{Schmidt,Honecker,Zhitomir,Derzhko2004,Zhit}. An interesting
and typical example of such flat-band system is the
$s=\frac{1}{2}$ delta or sawtooth Heisenberg model consisting of a
linear chain of triangles as shown in Fig.\ref{triangles}. The
interaction $J_{1}$ acts between the apical (even) and the basal
(odd) spins, while $J_{2}$ is the interaction between the
neighboring basal spins. The Hamiltonian of this model has a form
\begin{eqnarray}
H &=&J_{1}\sum_{i=1}^{N}\left(
s_{i}^{x}s_{i+1}^{x}+s_{i}^{y}s_{i+1}^{y}+\Delta _{1}(s_{i}^{z}s_{i+1}^{z}-%
\frac{1}{4})\right)  \nonumber \\
&&+J_{2}\sum_{i=1}^{N}\left(
s_{2i-1}^{x}s_{2i+1}^{x}+s_{2i-1}^{y}s_{2i+1}^{y}+\Delta
_{2}(s_{2i-1}^{z}s_{2i+1}^{z}-\frac{1}{4})\right) \label{1}
\end{eqnarray}%
where $s_{i}^{\lambda }$ are $s=\frac{1}{2}$ operators, $\Delta
_{1}$ and $\Delta _{2}$ are parameters representing the anisotropy
of basal-apical and basal-basal exchange interactions
respectively, $N$ is the number of sites. For the periodic
boundary conditions (PBC) $s_{1}=s_{N+1}$. The constants in
Eq.(\ref{1}) are chosen so that the energy of the ferromagnetic
state with $S^{z}=\pm \frac{N}{2}$ is zero.

The ground state of the isotropic Heisenberg model ($\Delta _{1}=$
$\Delta _{2}=1$) with both antiferromagnetic interactions
($J_{1}$, $J_{2}>0$) (AF-delta-chain) has been studied as a
function of parameter $\frac{J_{2}}{J_{1}}$
\cite{Sen,Nakamura,Blundell}. Remarkably, for a special choice of
the interaction values $J_{1}=2J_{2}$ the lower one-magnon band is
dispersionless and the excitations in this band are localized
states. The localized nature of the one- magnon states is a base
for the construction of multi-magnon states because a state
consisting of $k$ independent (non-overlapping) localized magnons
is an exact eigenstate. Such construction is possible up to $k\leq
\frac{N}{4}$ and these states form the ground state manifold at
the saturation magnetic field. The ground state and
low-temperature properties of the AF delta-chain with
$J_{1}=2J_{2} $ have been studied in detail in
Refs.\cite{Zhit,Schmidt,Derzhko2004,Honecker}. Typical features
related to the localized magnon states are the zero-temperature
magnetization plateau and the magnetization jump, residual entropy
and extra low-temperature peak in the specific heat.

In contrast to the AF delta-chain the same model with $J_{1}<0$
and $J_{2}>0$ (F-AF delta-chain) is less studied though this model
is used as a model of several compounds such as
$[Cu(bpy)H_{2}O]\times \lbrack Cu(bpy)(mal)H_{2}O](ClO_{4})_{2}$
containing magnetic $Cu^{2+}$ ions
\cite{Inagaki,Tonegawa,ruiz,Kaburagi}. Similarly to the AF
delta-chain in this model the localized states exist if
$\left\vert J_{1}\right\vert =2J_{2}$. It is known \cite{Tonegawa}
that the ground state of the F-AF isotropic delta chain is
ferromagnetic for $\alpha =\frac{J_{2}}{\left\vert
J_{1}\right\vert }<\frac{1}{2}$. In Ref.\cite{Tonegawa} it was
argued that the ground state for $\alpha >\frac{1}{2}$ is a
special ferrimagnetic state. The critical point $\alpha
=\frac{1}{2}$ is the transition point between these two ground
state phases. The isotropic F-AF delta-chain at the transition
point $\alpha =\frac{1}{2}$ has been studied in Ref.\cite{KDNDR}.
It was shown \cite{KDNDR} that in addition to the multi-magnon
configurations consisting of isolated magnons the special states
with overlapping magnons (localized multi-magnon complex) exist
and all of them are exact ground states at zero magnetic field.
So, the ground state degeneracy in this model is even larger than
for the AF delta-chain. Another difference between two isotropic
models concerns the energy gaps between the ground state and the
excited states. In the AF delta chain these gaps are finite while
in the F-AF model the gaps for $k$-magnon states decrease rapidly
with the increase of $k$. As a result the contribution of the
excited states to the thermodynamics can not be neglected.

It is interesting to study the influence of the anisotropy of
exchange interactions in the F-AF delta-chain on the ground state
properties and on the low-temperature thermodynamics. As it will
be shown there is a special line in ($\Delta _{1}$,$\Delta _{2}$)
plane on which the localized magnons are exact ground states in
zero magnetic field. The main aim of this paper is to study the
F-AF delta-chain on this line. We will demonstrate that the
behavior of the model on this line has non-trivial peculiarities

The paper is organized as follows. In Section II we derive the
conditions on model parameters that provides localized magnon
eigenstates and, therefore, macroscopic degeneracy of the ground
state, which is calculated in Section III. In Section IV we study
the low-temperature thermodynamic of the system both analytically
and numerically. In Section V we give a summary of our results.

\section{One-magnon states}

We begin the study of the anisotropic F-AF chain described by
Eq.(\ref{1}) with the one-magnon spectrum over ferromagnetic
state. Two branches of states with $S^{z}=S_{\max }^{z}-1$ are
given by
\begin{equation}
E_{\pm }(q)=\Delta _{1}-\frac{1}{2}%
\left[ \alpha (\Delta _{2}-\cos q)\pm \sqrt{\alpha
^{2}(\Delta_2-\cos q)^{2}+2(1+\cos q)}\right]  \label{2}
\end{equation}
here and further we put $J_{1}=-1$.

At a definite value of $\alpha=\alpha_0$ with:
\begin{equation}
\alpha _{0}=\frac{1}{\sqrt{2(1+\Delta _{2})}}  \label{3}
\end{equation}%
the lower band becomes dispersionless with the energy
\begin{equation}
\varepsilon=\Delta _{1}-\frac{1}{2\alpha _{0}}  \label{4}
\end{equation}

We note that the value of $\alpha _{0}$ does not depend on $\Delta
_{1}$ but the energy $\varepsilon $ does. The dispersionless
one-magnon states correspond to localized states which can be
chosen as
\begin{equation}
\hat{\varphi}_{i}\left\vert F\right\rangle =(s_{2i}^{-}+\frac{1}{\alpha _{0}}%
s_{2i+1}^{-}+s_{2i+2}^{-})\left\vert F\right\rangle \quad i=1,\ldots n
\label{5}
\end{equation}%
where $\left\vert F\right\rangle $ is the ferromagnetic state with
all spins up, $s_{i}^{-}$ is on-site spin lowering operator and
$n=N/2$.

The wave function $\hat{\varphi}_{i}\left\vert F\right\rangle $ is
localized in a valley between $i$-th and ($i+1$)-th triangles. The
wave functions Eq.(\ref{5}) are eigenfunctions of $\hat{H}$ at
$\alpha =\alpha _{0}$. To prove it let us represent the
Hamiltonian $\hat{H}$ as a sum of local Hamiltonians
\begin{equation}
\hat{H}=\sum_{i}\hat{H}_{i}  \label{6}
\end{equation}%
where $\hat{H}_{i}$ is the Hamiltonian of the $i$-th triangle
which is
\begin{eqnarray}
\hat{H}_{i} &=&-\sum_{\delta =\pm 1}\left( s_{2i+\delta
}^{x}s_{2i}^{x}+s_{2i+\delta }^{y}s_{2i}^{y}+\Delta _{1}(s_{2i+\delta
}^{z}s_{2i}^{z}-\frac{1}{4})\right)   \nonumber \\
&&+\alpha \left( s_{2i-1}^{x}s_{2i+1}^{x}+s_{2i-1}^{y}s_{2i+1}^{y}+\Delta
_{2}(s_{2i-1}^{z}s_{2i+1}^{z}-\frac{1}{4})\right)   \label{7}
\end{eqnarray}%

It is easy to check that at $\alpha =\alpha _{0}$%
\begin{equation}
(\hat{H}_{i}+\hat{H}_{i+1})\hat{\varphi}_{i}\left\vert
F\right\rangle =\varepsilon \hat{\varphi}_{i}\left\vert
F\right\rangle  \label{8}
\end{equation}%
and $\hat{H}_{j}\hat{\varphi}_{i}\left\vert F\right\rangle =0$ for
$j\neq i,i+1$.

Therefore,%
\begin{equation}
\hat{H}\hat{\varphi}_{i}\left\vert F\right\rangle =\varepsilon
\hat{\varphi}_{i}\left\vert F\right\rangle  \label{9}
\end{equation}

Because the one-magnon wave function Eq.(\ref{5}) is localized it
is possible to construct the states with $k$ independent
(non-overlapping) localized magnons for $k\leq \frac{N}{4}$ with
the energy $E_{k}=\varepsilon k$.

Thus we found that model Eq.(\ref{1}) for definite choice of
parameter $\alpha=\alpha_0$ given by Eq.(\ref{3}) has localized
magnon eigenstates Eq.(\ref{5}) with the energy $\varepsilon$
Eq.(\ref{4}). If parameters $\Delta_1$ and $\Delta_2$ are chosen
so that $\varepsilon <0$ ($\Delta _{2}>2\Delta _{1}^{2}-1$), all
the states composed of $k\leq \frac{N}{4}$ independent localized
magnons are the lowest ones in the corresponding spin sector
$S^z=\frac{N}{2}-k$. It turns out that the lowest state in the
case $\varepsilon <0$ lies in the sector $S^z=0$
\cite{Derzhko2008}. The magnetic properties of the F-AF
delta-chain at $\alpha =\alpha _{0}$ and $\varepsilon <0$ are
similar to those for the AF isotropic delta-chain with
$J_{1}=2J_{2}$. In particular, the ground state magnetization
curve has the plateau and the jump, and all the states composed of
independent localized magnons form the macroscopically degenerated
ground state at the saturation magnetic field.

If $\varepsilon >0$ ($\Delta _{2}<2\Delta _{1}^{2}-1$) the
localized magnons are exact eigenstates as well, but they are not
the lowest eigenstates. Moreover, if $\varepsilon >0$ the ground
state is ferromagnetic with $S_{total}^{z}=\pm \frac{N}{2}$. To
prove it let us consider eigenvalues of the triangle Hamiltonian
with $\alpha =\alpha _{0}$. The spectrum of each local
$\hat{H}_{i}$ consists of four levels, all of them are two-fold
degenerated over $S_{z}$:
\begin{eqnarray}
E_{1} &=&0,\quad S_{z}=\pm \frac{3}{2}  \nonumber \\
E_{2} &=&\frac{\varepsilon }{2},\quad S_{z}=\pm \frac{1}{2}
\nonumber \\
E_{3,4} &=&\frac{1}{2}(\alpha _{0}+\frac{3\varepsilon
}{2}+\frac{1}{2\alpha _{0}})\pm \frac{1}{2}\sqrt{(\alpha
_{0}+\frac{\varepsilon }{2}+\frac{1}{2\alpha _{0}})^{2}-2\alpha
_{0}\varepsilon },\quad S_{z}=\pm \frac{1}{2} \label{10}
\end{eqnarray}

It follows from Eq.(\ref{10}) that all eigenvalues for $S_{z}=\pm
\frac{1}{2} $ in Eq.(\ref{10}) are positive at $\varepsilon >0$.
Since the local Hamiltonians of neighbor triangles do not commute
with each other the ground state energy $E_{0}$ of $\hat{H}$
satisfies an inequality
\begin{equation}
E_{0}\geq \sum_{i}E_{0i}=0  \label{11}
\end{equation}%
where $E_{0i}=0$ is the ground state energy of $i$-th triangle
with $S_{z}=\pm \frac{3}{2}$. The inequality (\ref{11}) turns in
an equality only if all triangles have $S_{z}=\frac{3}{2}$ or
$S_{z}=-\frac{3}{2}$ simultaneously. Then, the states with
$S_{total}^{z}=\pm \frac{N}{2}$ are two ground states only.

Thus, we conclude that the ground state with $S_{total}^{z}=0$ is
realized if $\varepsilon <0$ while for $\varepsilon >0$ the ground
state is ferromagnetic. However, if $\varepsilon =0$ the lowest
eigenvalues with $S_{z}=\pm \frac{1}{2}$ of the local Hamiltonian
becomes zero and it indicates the possibility of a significant
increase of the ground state degeneracy.

The condition $\varepsilon =0$ means that model parameters
$\Delta_1$ and $\Delta_2$ are coupled and we parameterize them by
the anisotropy of the basal-apical interaction $\Delta_1$:
\begin{equation}
\Delta_{2} = 2\Delta_{1}^{2}-1  \label{12}
\end{equation}

Hamiltonian (\ref{1}) of the anisotropic F-AF delta-chain with
this choice and at $\alpha=\alpha_0$ has a form
\begin{eqnarray}
H &=&-\sum \left( s_{i}^{x}s_{i+1}^{x}+s_{i}^{y}s_{i+1}^{y}+
\Delta_1\left( s_{i}^{z}s_{i+1}^{z}-\frac{1}{4}\right) \right)
\nonumber \\ &&+\frac{1}{2\Delta_1} \sum \left(
s_{2i-1}^{x}s_{2i+1}^{x}+s_{2i-1}^{y}s_{2i+1}^{y}+(2\Delta
_{1}^{2}-1)\left( s_{2i-1}^{z}s_{2i+1}^{z}-\frac{1}{4}\right)
\right) \label{13}
\end{eqnarray}%

Model (\ref{13}) is the main object of the following study. It has
macroscopic degeneracy of the ground state and nontrivial
thermodynamic properties. When the parameter $\Delta_1$ is changed
from $\Delta_1=0$ to $\Delta_1=\infty $ the model interpolates
between the isotropic ferromagnetic chain on the basal sites and
the Ising model with equal but opposite in sign basal-apical and
basal-basal interactions. In the case $\Delta_1=1$ model
(\ref{13}) reduces to the isotropic F-AF delta-chain studied in
Ref.\cite{KDNDR}.

As noted in Ref.\cite{KDNDR,Tonegawa}, Hamiltonian (\ref{13}) with
$\Delta_1=1$ describes the model at the transition point between
the ferromagnetic and the ferrimagnetic ground states. In this
respect model (\ref{13}) as a function of $\Delta_1$ describes the
transition line of more general model, where the parameter
$\alpha$ is not put as $\alpha=\frac{1}{2\Delta_1}$. This model
reads:
\begin{eqnarray}
H &=&-\sum \left( s_{i}^{x}s_{i+1}^{x}+s_{i}^{y}s_{i+1}^{y}+\Delta
_{1}\left( s_{i}^{z}s_{i+1}^{z}-\frac{1}{4}\right) \right)
\nonumber \\ &&+\alpha \sum \left(
s_{2i-1}^{x}s_{2i+1}^{x}+s_{2i-1}^{y}s_{2i+1}^{y}+(2\Delta
_{1}^{2}-1)\left( s_{2i-1}^{z}s_{2i+1}^{z}-\frac{1}{4}\right)
\right) \label{14}
\end{eqnarray}%

\begin{figure}[tbp]
\includegraphics[width=3in,angle=-90]{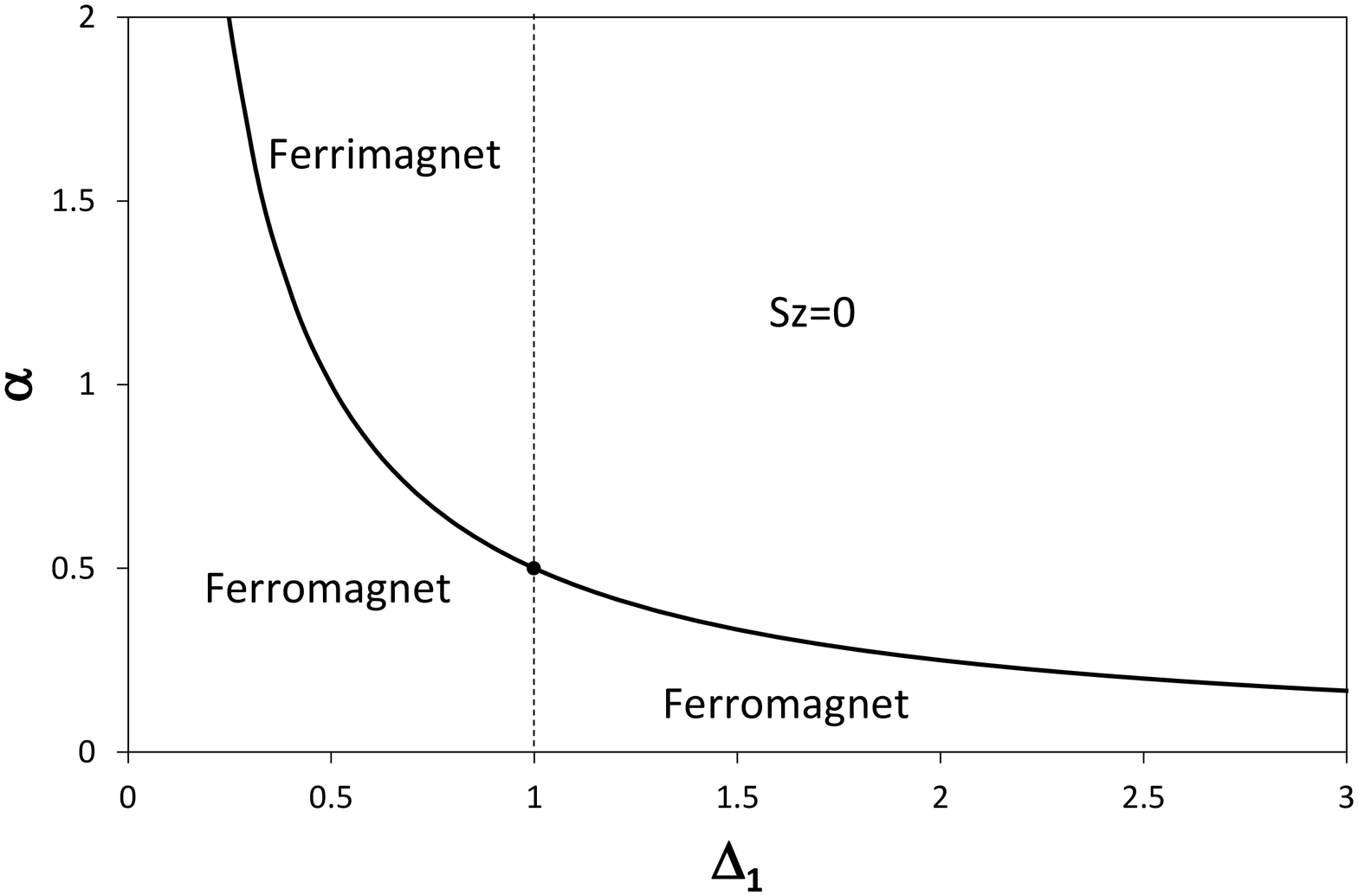}
\caption{The ground state phase diagram of model (\ref{14}).}
\label{phases}
\end{figure}

The ground state phase diagram of model (\ref{14}) in
($\Delta_1,\alpha $) plane obtained by numerical calculations is
shown in Fig.\ref{phases}. The curve $\alpha =\frac{1}{2\Delta_1}$
corresponds to the transition line between different phases shown
in Fig.\ref{phases}. We note that the study of the behavior of the
model in these phases is out of scope of the present paper and
will be given elsewhere. Here we focus on Hamiltonian (\ref{13}).

\section{The ground state degeneracy}

In this Section we study the ground state of Hamiltonian
(\ref{13}).

As follows from Eq.(\ref{11}) the ground state energy of this
Hamiltonian is zero. There are $n$ one-magnon states Eq.(\ref{5})
in the spin sector $S_{total}^{z}=\frac{N}{2}-1$. These states
form a complete nonorthogonal basis, have zero energy and,
therefore, belong to the ground state manifold.

It is evident that the pairs of the isolated magnons $\varphi
_{i}\varphi _{j}$ ($\left\vert i-j\right\vert >1$) are the
eigenfunctions with zero energy of each local Hamiltonian (and,
therefore, of total Hamiltonian) in the spin sector
$S_{total}^{z}=\frac{N}{2}-2$. Similarly, eigenfunction composed
of $k$ isolated magnons is the exact eigenfunction of the ground
state in the sector $S_{total}^{z}=\frac{N}{2}-k$. However, such
set of functions contains not all of the states with zero energy
in $k$-magnon sector with $k>1$. As it was shown in
Ref.\cite{KDNDR} for the isotropic F-AF delta-chain ($\Delta
_{1}=1$) there are also the ground state eigenfunctions of another
type, and this type of functions holds in model (\ref{13}) with
$0<\Delta _{1}<\infty $ as well. As an example, we consider
two-magnon eigenstates. For $k=2$ along with the pair of isolated
magnons we can write the exact two-magnon state as
\begin{equation}
\hat{\varphi}_{i}(\hat{\varphi}_{i-1}+B\hat{\varphi}_{i}
+\hat{\varphi}_{i+1})\left\vert F\right\rangle   \label{15}
\end{equation}%
where $B=$ $2\Delta _{1}^{2}-1$.

It is easy to check that function (\ref{15}) is the exact
eigenfunction with zero energy for the local Hamiltonians
$\hat{H}_{i}$, $\hat{H}_{i+1}$ and $\hat{H}_{i-1}$ and for the
others ones. For $\Delta_1=1$ this function reduces to that
considered in Ref.\cite{KDNDR}. We note that the function
(\ref{15}) contains overlapping magnons. The complete manifold of
ground states in the sector $S_{total}^{z}=\frac{N}{2}-2$ consists
of $\frac{n(n-3)}{2}$ pairs ($n=N/2$ is assumed to be even) of
independent magnons and $n$ eigenfunctions (\ref{15}). Thus, the
ground state degeneracy in this spin sector is $C_{n}^{2}$.

The construction of the wave functions of type Eq.(\ref{15}) can
be extended for $k>2$. According to the results of
Ref.\cite{KDNDR} the total number of the ground states of the
isotropic F-AF chain for fixed $S_{total}^{z}=(n-k)$ ($0\leq k\leq
n$) is
\begin{eqnarray}
G_{n}(k) &=&C_{n}^{k},\qquad 0\leq k\leq \frac{n}{2}  \nonumber \\
G_{n}(k) &=&C_{n}^{n/2}+\delta_{k,n},\qquad \frac{n}{2}\leq k\leq
n \label{17}
\end{eqnarray}

The total degeneracy of the ground state $W$ is
\begin{eqnarray}
W &=&2^{n}+nC_{n}^{n/2}+1  \label{18}
\end{eqnarray}
so that $W=2^{n}\sqrt{\frac{2n}{\pi }}$ in the limit $N\to\infty$.

The analysis similar to that for the isotropic model can be
carried out for the anisotropic delta-chain Eq.(\ref{13}). It
turns out that the degeneracy of the ground state given by
Eqs.(\ref{17})-(\ref{18}) is valid for all $0<\Delta _{1}<\infty $
except two special cases $\Delta_1=1/\sqrt{2}$ and $\Delta_1=1/2$
which are considered below.

In the case $\Delta_1=1/\sqrt{2}$ the anisotropy $\Delta_2$
vanishes and the model reduces to
\begin{equation}
H=-\sum \left(
s_{i}^{x}s_{i+1}^{x}+s_{i}^{y}s_{i+1}^{y}+\frac{1}{\sqrt{2}}
s_{i}^{z}s_{i+1}^{z} \right) +\frac{1}{\sqrt{2}} \sum \left(
s_{2i-1}^{x}s_{2i+1}^{x}+s_{2i-1}^{y}s_{2i+1}^{y}\right)
\label{19}
\end{equation}

In this special case the states containing neighbor localized
magnons like $\hat{\varphi}_{i}\hat{\varphi}_{i+1}$ are exact
ground states ($B=0$ in Eq.(\ref{15})), which is not valid in
general case. We note that similar type of exact eigenstates
arises in the Hubbard model on the delta-chain in which
neighboring valleys can be occupied by the localized electrons
with identical spins \cite{hub1,hub2}. The presence of such exact
states for model Eq.(\ref{19}) leads to the increase of the ground
state degeneracy, which is
\begin{equation}
W =2^{n}\left( \frac{n}{2}+1\right) ,\qquad
\Delta_1=\frac{1}{\sqrt{2}} \label{20}
\end{equation}

In the case $\Delta _{1}=1/2$ after rotation in the XY plane
$s_{m}^{x,y}\to (-1)^{m}s_{m}^{x,y}$ the Hamiltonian takes the
form
\begin{equation}
H=\sum \left(
s_{i}^{x}s_{i+1}^{x}+s_{i}^{y}s_{i+1}^{y}-\frac{1}{2}
s_{i}^{z}s_{i+1}^{z}\right) +\sum \left(
s_{2i-1}^{x}s_{2i+1}^{x}+s_{2i-1}^{y}s_{2i+1}^{y}-\frac{1}{2}
s_{2i-1}^{z}s_{2i+1}^{z}\right)  \label{21}
\end{equation}

That is basal-apical and basal-basal interactions are exactly
equal in this case. This fact raises symmetry of the system:
isosceles triangles in general case becomes equilateral triangles
in this special case. The increased symmetry results in the
additional degeneracy of the ground state, which is
\begin{equation}
W=2^{n}\left( \frac{n}{3}+1\right) +\frac{2n}{3} +1 ,\qquad
\Delta_1=\frac{1}{2} \label{22}
\end{equation}

In the limit $\Delta _{1}=0$ the model reduces to the quantum
ferromagnet on basal spins and independent $n$ apical spins.
Therefore, the ground state degeneracy is
\begin{equation}
W = \left( n+1\right) 2^{n} ,\qquad \Delta_1=0 \label{23}
\end{equation}
where the factor $(n+1)$ comes from the degeneracy of the
ferromagnetic state over total $S^{z}$.

The special case $\Delta_1\to\infty$ will be studied in detail in
Sec.IV. Here we give only the results for the ground state
degeneracy: $W=3^{n}+1$.

It is interesting to note that the ground state degeneracy of the
anisotropic delta-chain with the open boundary conditions (OBC) is
the same as for the isotropic model for all value of $0<\Delta
_{1}<\infty$. According to the results of Ref.\cite{KDNDR} the
ground state degeneracy of the open chains with add $N$ sites is
\begin{equation}
W_{OBC}=2^{n-1}(n+1), \qquad n=\frac{N+1}{2} \label{24}
\end{equation}

All the above presented expressions for the ground state
degeneracy have been confirmed by exact diagonalization (ED)
calculations of finite chains.

The exponential degeneracy of the ground state results in the
residual entropy $s_{0}=\ln (W)/N$. Though the numbers of
degenerated states in the general case Eq.(\ref{18}) and in
special cases Eqs.(\ref{20}),(\ref{22}),(\ref{23}),(\ref{24}) are
different, they yield the same result for the residual entropy in
the thermodynamic limit $N\to\infty$:
\begin{equation}
s_{0}=\frac{1}{2}\ln 2  \label{25}
\end{equation}

The difference in the numbers of degenerated ground states reveals
itself in the corrections $\sim \ln(N)/N$, that vanishes in the
thermodynamic limit.

\section{Low-temperature thermodynamics}

\subsection{Ising model}

In this Section we study the low-temperature behavior of the
anisotropic F-AF delta-chain on the transition line. We begin with
a limit $\Delta_1 \to \infty $ when model (\ref{13}) reduces to
the Ising model, the Hamiltonian of which has a form:
\begin{equation}
\hat{H}_{I}=-\sum_{i}(s_{i}^{z}s_{i+1}^{z}-\frac{1}{4})+
\sum_{i}(s_{2i-1}^{z}s_{2i+1}^{z}-\frac{1}{4})-h\sum_{i}s_{i}^{z}
\label{27}
\end{equation}
where we introduced the dimensionless magnetic field $h$.

The partition function of this model can be obtained using a
transfer-matrix method and given by
\begin{equation}
Z=\lambda _{1}^{n}+\lambda _{2}^{n}  \label{28}
\end{equation}%
where eigenvalues of the transfer-matrix are
\begin{eqnarray}
\lambda _{1} &=&1+2\cosh \left( \frac{h}{T}\right) +\exp \left( -\frac{1}{T}%
\right)  \nonumber \\
\lambda _{2} &=&-1+\exp \left( -\frac{1}{T}\right) \quad  \label{29}
\end{eqnarray}%

The ground state of model (\ref{27}) at $h=0$ has zero energy and
the ground state degeneracy $G_{n}(k)$ in the spin sector
$S_{total}^{z}=(n-k)$ ($0\leq k\leq n$) can be found as
coefficients in the expansion of $Z$ in powers of $\exp(-h/T)$. As
a result $G_{n}(k)$ is given by
\begin{equation}
G_{n}(k)=\sum_{m=0}^{k}C_{n}^{m+n-k}C_{m+n-k}^{m/2}+\delta _{k,n}
\label{30}
\end{equation}

The total ground state degeneracy is%
\begin{equation}
W=2\sum_{k=0}^{n-1}G_{n}(k)+G_{n}(n)=3^{n}+1  \label{31}
\end{equation}

The residual entropy per site $s_{0}$ of the Ising model
(\ref{27}) at $N\to \infty $ equals
\begin{equation}
s_{0}=\frac{1}{2}\ln 3  \label{32}
\end{equation}

Using Eqs.(\ref{28}) and (\ref{29}) we can obtain all
thermodynamic quantities (in the thermodynamic limit only largest
eigenvalue $\lambda _{1}$ survives). In particular, the specific
heat per site is given by
\begin{equation}
C=\frac{3}{2T^{2}\left[ 3\exp \left( \frac{1}{2T}\right) +\exp
(-\frac{1}{2T})\right] ^{2}}  \label{33}
\end{equation}

The specific heat as a function of temperature has a typical broad
maximum around $T\simeq 0.5$ and exponential decay for $T\to 0$.
The zero-field susceptibility per site $\chi $ is given by
\begin{equation}
\chi =\frac{1}{T\left[ 3+\exp \left( -\frac{1}{T}\right) \right] }
\label{34}
\end{equation}
and it behaves as $\chi \sim \frac{1}{3T}$ for $T\to 0$.

Now let us consider the generalization of the Ising model
(\ref{27}) where basal-apical and basal-basal interactions are
different:
\begin{equation}
\hat{H}_I = -\sum_{i}(s_{i}^{z}s_{i+1}^{z}-\frac{1}{4})+(1+\gamma)
\sum_{i}(s_{2i-1}^{z}s_{2i+1}^{z}-\frac{1}{4})-h\sum_{i}s_{i}^{z}
\label{26}
\end{equation}%

The ground state of model (\ref{26}) is ferromagnetic for
$\gamma<0$ and the `antiferromagnetic' with $S^{z}=0$ on the basal
subsystem for $\gamma>0$. The Ising model (\ref{27}) ($\gamma=0$)
describes the transition point between these phases. Generally
speaking, the Ising model (\ref{26}) with $\gamma\neq 0$ is not
any limiting case of the initial model (\ref{13}). Nevertheless,
it is useful to study model (\ref{26}) with $\gamma>0$ because on
the one hand it has exact solution and on the other hand some
properties of the thermodynamics in the AF phase, especially in
the vicinity of the transition point, are inherent in model
(\ref{13}) as well.

The eigenvalues of the transfer-matrix are
\begin{equation}
\lambda _{1,2} = \cosh \left( \frac{h}{T}\right) +\exp\left(
-\frac{1}{T}\right) \pm \sqrt{\left[ 1+\cosh \left(
\frac{h}{T}\right) \right] \left[ 2\exp \left(\frac{\gamma
}{T}\right)+\cosh \left( \frac{h}{T}\right) -1\right] } \label{35}
\end{equation}

The partition function at $N\gg 1$ is%
\begin{equation}
Z=2^{n}\exp \left( \frac{\gamma n}{2T}\right) \left[
1+\frac{1}{2}\exp \left( -\frac{\gamma }{2T}\right)
+\frac{1}{2}\exp \left( -\frac{2+\gamma }{2T}\right) \right] ^{n}
\label{36}
\end{equation}

The ground state degeneracy $W=3^{n}$ for $\gamma =0$ is partially
lifted up to $2^{n}$ for $\gamma >0$. The degeneracy $2^{n}$ is
related to an independence of the ground state energy for $\gamma
>0$ on the spin configuration of the apical subsystem.

The temperature dependence of the specific heat obtained from
Eq.(\ref{36}) has a form
\begin{equation}
C(T)=\frac{(2+\gamma )^{2}\exp \left( -\frac{2+\gamma }{2T}\right)
+\gamma ^{2}\exp \left( -\frac{\gamma }{2T}\right) +2\exp \left(
-\frac{1+\gamma }{T}\right) }{4T^{2}\left[ 1+2\exp \left(
-\frac{\gamma }{2T}\right) +\exp \left( -\frac{2+\gamma
}{2T}\right) \right] ^{2}}  \label{37}
\end{equation}

\begin{figure}[tbp]
\includegraphics[width=3in,angle=-90]{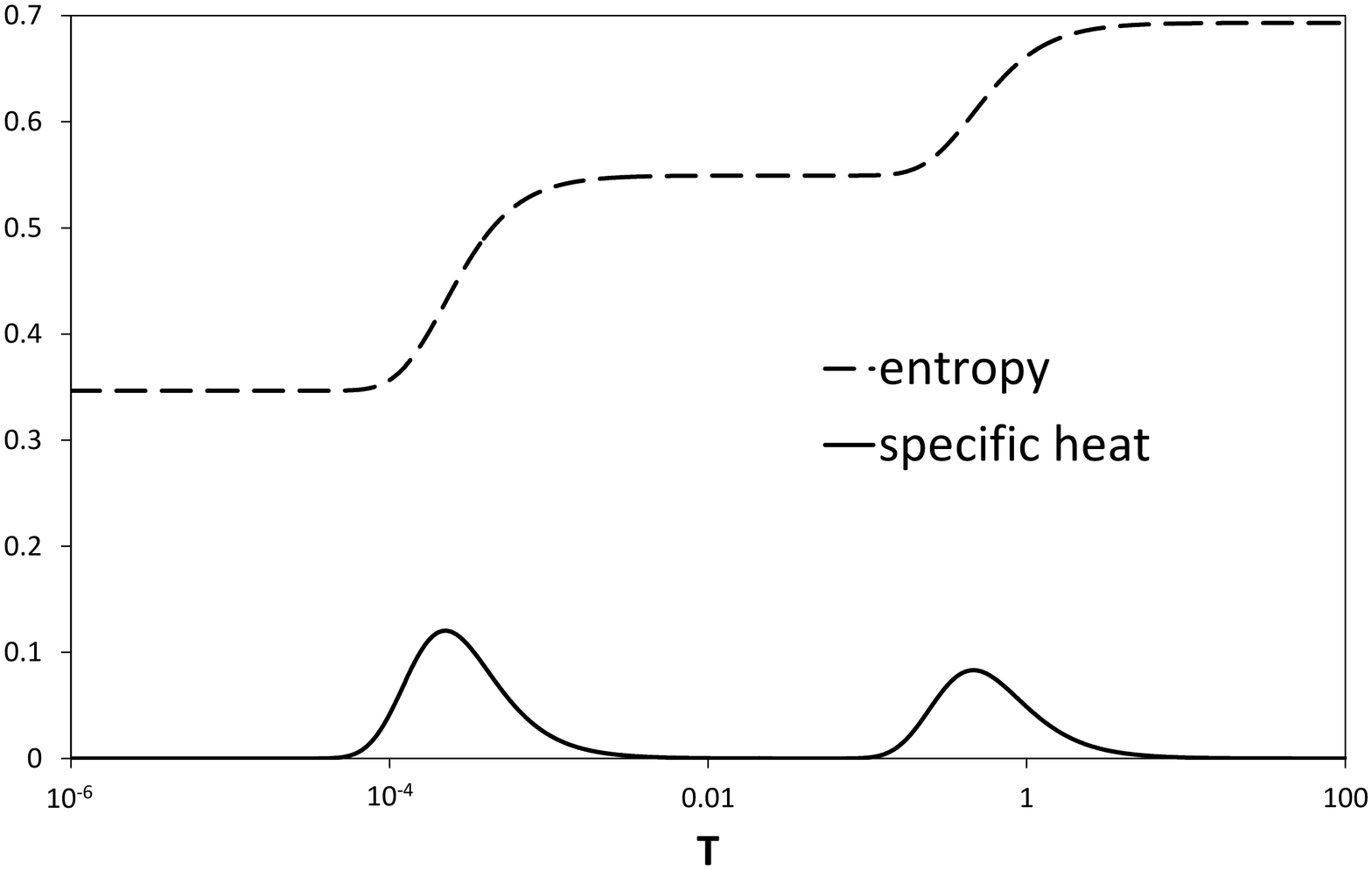}
\caption{The specific heat and entropy as the function of $T$ of
model (\ref{26}) for $\gamma =0.001$.} \label{ising}
\end{figure}

The specific heat as a function of $T$ is shown in Fig.\ref{ising}
for $\gamma =0.001$. In comparison with the case $\gamma =0$ the
dependence $C(T)$ has an additional low-temperature maximum at
$T\simeq \frac{\gamma }{4}$. Such two-peak form of the temperature
dependence of the specific heat can be explained as follows. The
spectrum of the Ising model (\ref{26}) with $0<\gamma \ll 1$ has
two energy scales. One of them is set by ($4^n-3^n$) states with
energies $E\sim 1$ and another one is set by ($3^n-2^n$)
low-energy states that split-off from the ground state at small
$\gamma $. According to Eq.(\ref{36}) the density of the
low-energy states having the energy $E=k\gamma/2$ is $\rho
(k)=2^{n-k}C_{n}^{k}$ with a maximum at $k=\frac{n}{3}$. This part
of the energy spectrum is responsible for the appearance of the
low-temperature maximum in $C(T)$. The two-scale form of the
low-energy spectrum causes the peculiarities of the other
thermodynamic quantities. For example, the entropy per site $s(T)$
has stair-step-like temperature dependence (see Fig.\ref{ising}).
In the regions $T\ll \gamma$, $\gamma \ll T\ll 1$, $1\ll T$ the
entropy per site $s(T)$ behaves as $s\simeq \frac{1}{2}\ln 2$,
$s\simeq \frac{1}{2}\ln 3$, $s\simeq \frac{1}{2}\ln 4$,
correspondingly. The quantity $\chi T$ has similar
stair-step-likes behavior.

\subsection{F-AF delta-chain with $\Delta _{1}\gg 1$}

When the parameter $\Delta_1$ is large it is convenient to
normalize Hamiltonian (\ref{13}) as $\hat{H}/\Delta_1$ and to
write it in the form
\begin{equation}
\frac{1}{\Delta_1}\hat{H}=\hat{H}_{I}+\hat{V}_{1}+\hat{V}_{2}
\label{38}
\end{equation}%
where $\hat{H}_{I}$ is the Ising Hamiltonian (\ref{27}) at $h=0$
and $\hat{V}_{1}$ and $\hat{V}_{2}$ have the forms
\begin{eqnarray}
\hat{V}_{1} &=&-2g\sum (s_{i}^{x}s_{i+1}^{x}+s_{i}^{y}s_{i+1}^{y})
\nonumber \\
\hat{V}_{2} &=&2g^{2}\sum
(s_{2i-1}^{x}s_{2i+1}^{x}+s_{2i+1}^{y}s_{2i-1}^{y}
-s_{2i-1}^{z}s_{2i+1}^{z}+\frac{1}{4})  \label{39}
\end{eqnarray}%
with small parameter $g=1/(2\Delta _{1})$.

At $g=0$ the ground state of Hamiltonian (\ref{38}) is
$3^{n}$-fold degenerated and there are ($4^n-3^n$) states with
$E\geq 1$. The terms $\hat{V}_{1}$ and $\hat{V}_{2}$ lift the
degeneracy for each spin sector, but partly only: the part of the
ground state levels remains degenerated with zero energy (the
number of them is given by Eq.(\ref{18})) and other ones move up.
As a result the ground state degeneracy of model (\ref{38}) is
$W\simeq 2^{n}$ in comparison with $W_{I}=3^{n}$ at $g=0$. At
$g\ll 1$ the levels of Hamiltonian (\ref{38}) which split off from
the ground state form a set of low-lying excitations determining
the low-temperature thermodynamics.

The calculation of the spectrum of these states is very
complicated problem and we begin with the one-magnon subspace
$S_{total}^{z}=\frac{N}{2}-1$. In the one-magnon subspace the
number of degenerate ground states of model (\ref{38}) is $n$ for
both cases $g=0$ and $g\neq 0$, i.e. the perturbation
$\hat{V}=\hat{V}_{1}+\hat{V}_{2}$ does not lift the degeneracy. In
the two-magnon sector the ground state degeneracies of
$\hat{H}_{I}$ and of $\hat{H}$ are $C_{n+1}^{2}$ and $C_{n}^{2}$,
correspondingly. It means that $n$ states split off from the
ground state. In fact, these states are two-magnon bound complexes
and their energy found in the second order in $g$ is $E=g^{2}$.

The calculation of the PT in $g$ for $k=3$ is rather cumbersome
and we give the final result only. The set of $(C_{n}^{3}+n(n-1))$
three-magnon states which are degenerated at $g=0$ splits into
three subsets. One of them contains $C_{n}^{3}$ ground states with
$E=0$. There are $n(n-2)$ states with $E\sim g^{2}$. These states
consist of the two-magnon bound complex and the isolated localized
magnon. The third subset represents $n$ three-magnon bound
complexes with the energy $E=3g^{4}$.

The analytical computation of the spectrum of low-lying
excitations for $k>3$ is complicated problem and our further
conclusions are based on numerics. According to numerical data the
structure of the spectrum of low-lying $k$-magnon states is
similar to that for the case $k=3$. The perturbation
$\hat{V}=\hat{V}_{1}+\hat{V}_{2}$ splits the ground state manifold
into $k$-subsets: the ground states with the energies $E=0$;
$k$-magnon bound complexes with $E\sim g^{2(k-1)}$; the states
consisting of $(k-1)$-magnon bound complex and one isolated magnon
($E\sim g^{2(k-2)}$); the states consisting of $(k-2)$-magnon
bound complex and two isolated magnons ($E\sim g^{2(k-3)}$); and
so on. The highest subset of excitations has the energies $E\sim
g^{2}$. Thus, the low-lying excitations in the sector with
$S_{total}^{z}=\frac{N}{2}-k$ is distinctly divided into the parts
with the energies $E\sim g^{2},$ $E\sim g^{4},\ldots E\sim
g^{2(k-1)}$.

Taking into account all states with all possible values of
$S_{total}^{z}$, we found that the total spectrum of model
(\ref{38}) can be rank-ordered in powers of small parameter
$g^{2}$ and it has a multi-scale structure. The ground state
degeneracy is given by Eq.(\ref{18}). The lowest excited states
for finite chain composed of $n$ triangles has the energy $E\sim
g^{2n-2}$ and it means that the gap is exponentially small.

\begin{figure}[tbp]
\includegraphics[width=3in,angle=-90]{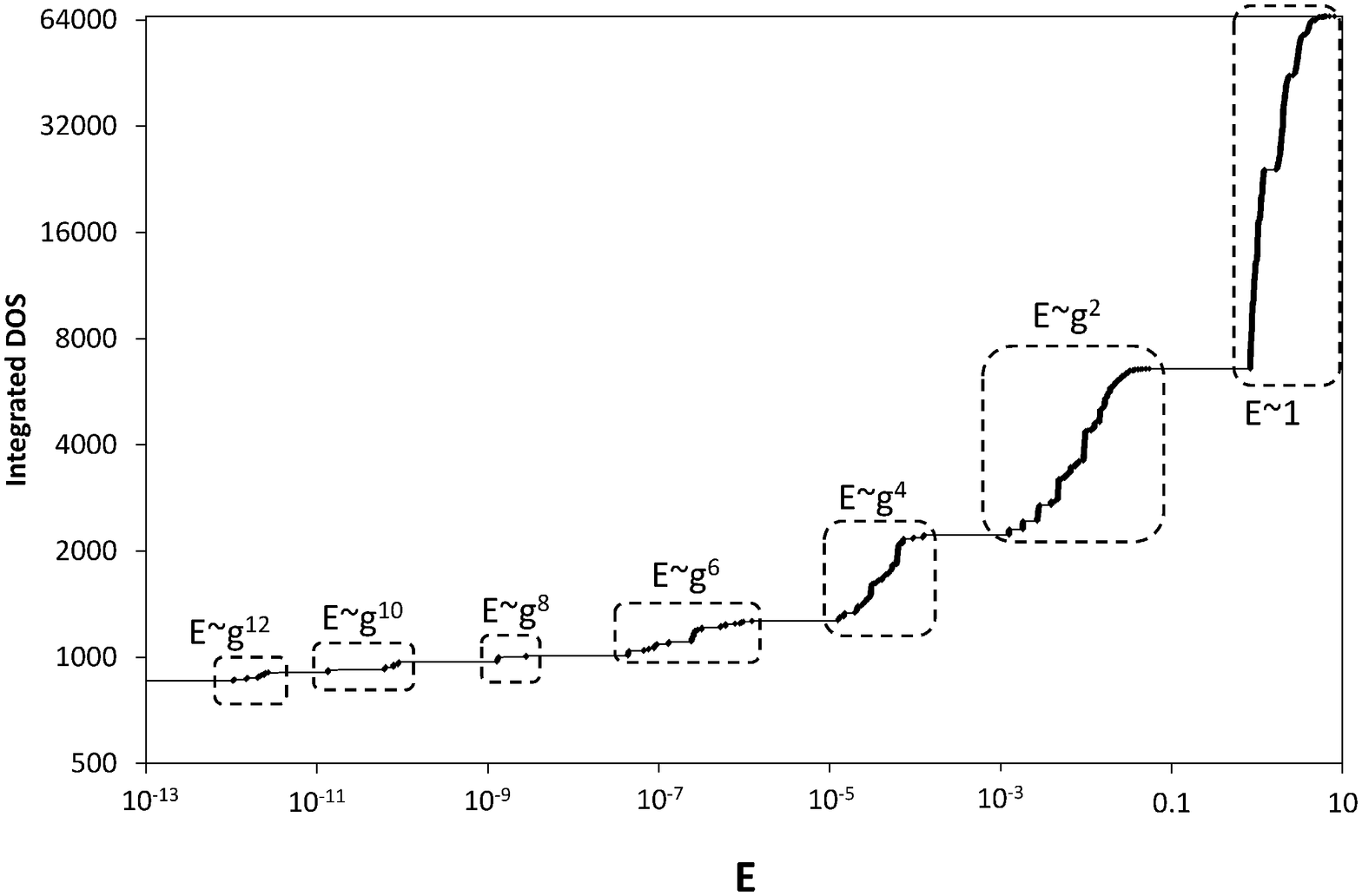}
\caption{The integrated density of states of normalized
Hamiltonian (\ref{38}) for $g=0.1$ and $N=16$.} \label{DOS}
\end{figure}

The distribution of the energy levels for $N=16$ and $g=0.1$ is
shown in Fig.\ref{DOS}. As it is seen in Fig.\ref{DOS} the
spectrum is distinctly divided into the parts. Each part of the
spectrum behaves as $E\sim g^{2k}$ as written in Fig.\ref{DOS}.
This fact was confirmed numerically by comparison of the energies
for $g=0.1$ and $g=0.125$. We note that for $N=16$ and $g=0.1$ the
value of the smallest excitation $E\sim g^{14}$ becomes very small
($E\sim 10^{-13} - 10^{-14}$) and it is indistinguishable from the
ground state, so only the excitations with $E>10^{-12}$ are
accessible by the ED calculations and represented in
Fig.\ref{DOS}.

\begin{figure}[tbp]
\includegraphics[width=3in,angle=-90]{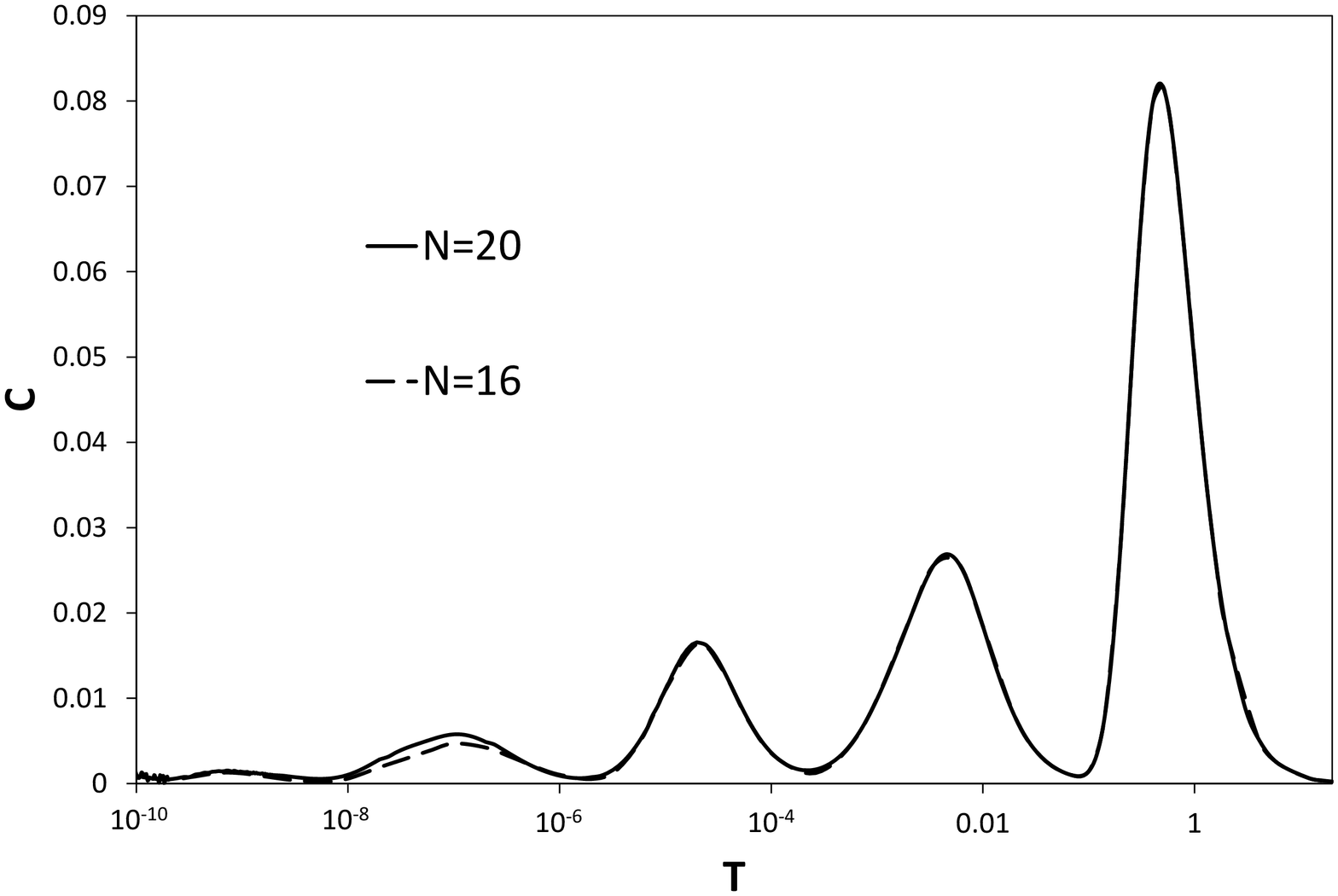}
\caption{The dependence of specific heat on the temperature of
normalized Hamiltonian (\ref{38}) for $g=0.1$ and $N=16,20$.}
\label{C_A10}
\end{figure}

\begin{figure}[tbp]
\includegraphics[width=3in,angle=-90]{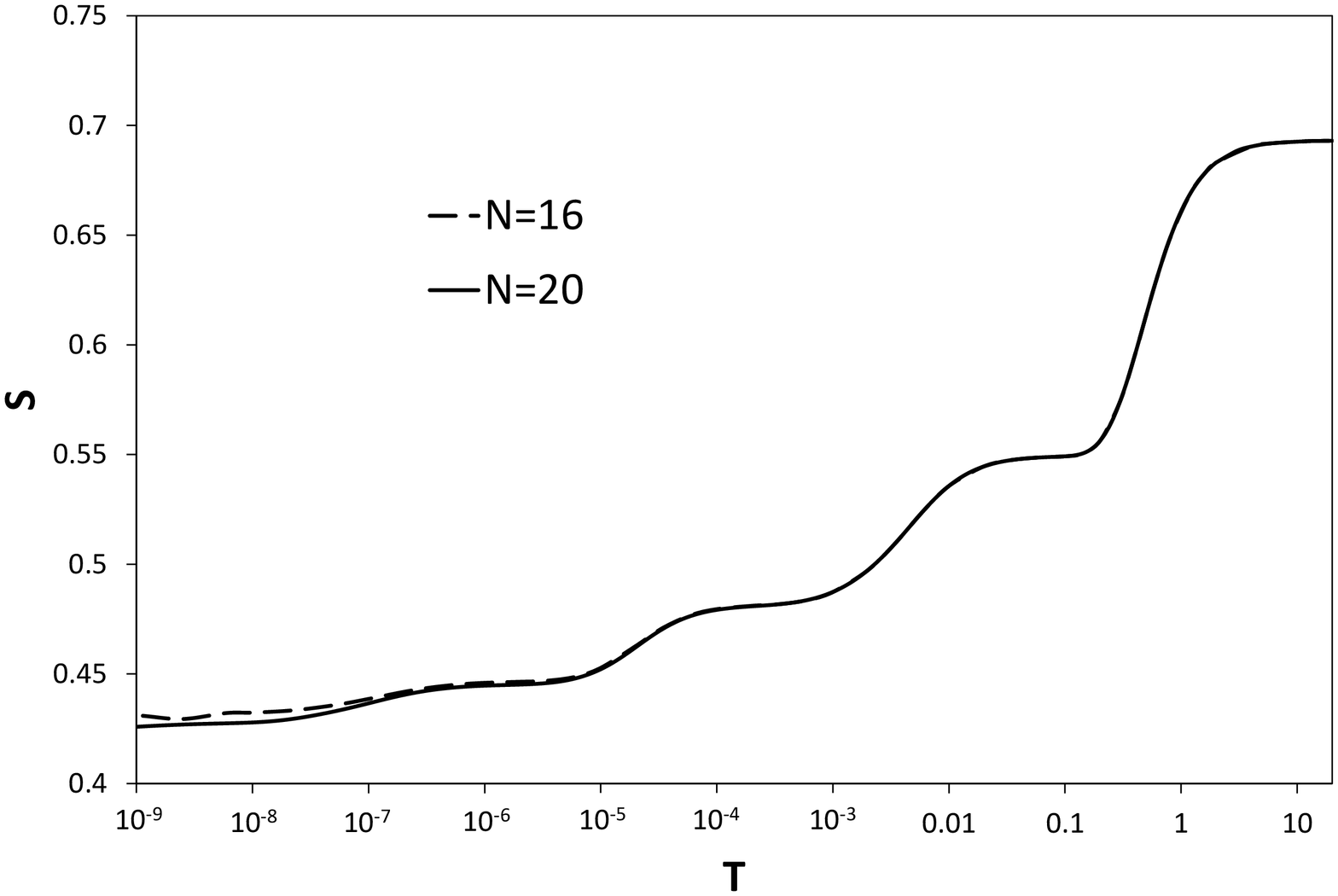}
\caption{The dependence of entropy on the temperature of
normalized Hamiltonian (\ref{38}) for $g=0.1$ and $N=16,20$.}
\label{S_A10}
\end{figure}

Such structure of the spectrum determines the characteristic
features of the low-temperature thermodynamics. To study the
thermodynamics of model (\ref{13}) we use the exact
diagonalization (ED) of finite delta-chains with PBC up to $N=20$.
In Figs.\ref{C_A10} and \ref{S_A10} we represent the data for the
specific heat $C(T)$ and the entropy $s(T)$ (both per site) for
$\Delta_1=5$ ($g=0.1$) obtained by the ED for $N=16$ and $N=20$.
The temperature dependence of the specific heat shown in
Fig.\ref{C_A10} exhibits numerous maxima the presence of which can
be explained as follows. As it was shown in Sec.IVa the dependence
$C(T)$ of the Ising model has one low-temperature maximum which is
related to the part of the spectrum with the energy $E\sim
\gamma^{2}$. Similarly, multi-scale structure of the spectrum of
Eq.(\ref{38}) for small $g$ leads to the dependence $C(T)$ with
many maxima related to the corresponding parts of the spectrum.
This is confirmed by the following calculation. Let us select the
part of the spectrum with $E\sim g^{2}$, remove the high-energy
part with $E\sim 1$, put the energy of all remaining low-lying
states to zero (as the ground state) and calculate the
contribution of such deformed spectrum to the specific heat. It
turns out that this contribution perfectly describes the first
low-temperature peak in $C(T)$. Similar procedure for the parts of
the spectrum with $E\sim g^{4}$, $E\sim g^{6}$ and so on
reproduces very well corresponding peaks in $C(T)$.

\begin{figure}[tbp]
\includegraphics[width=3in,angle=-90]{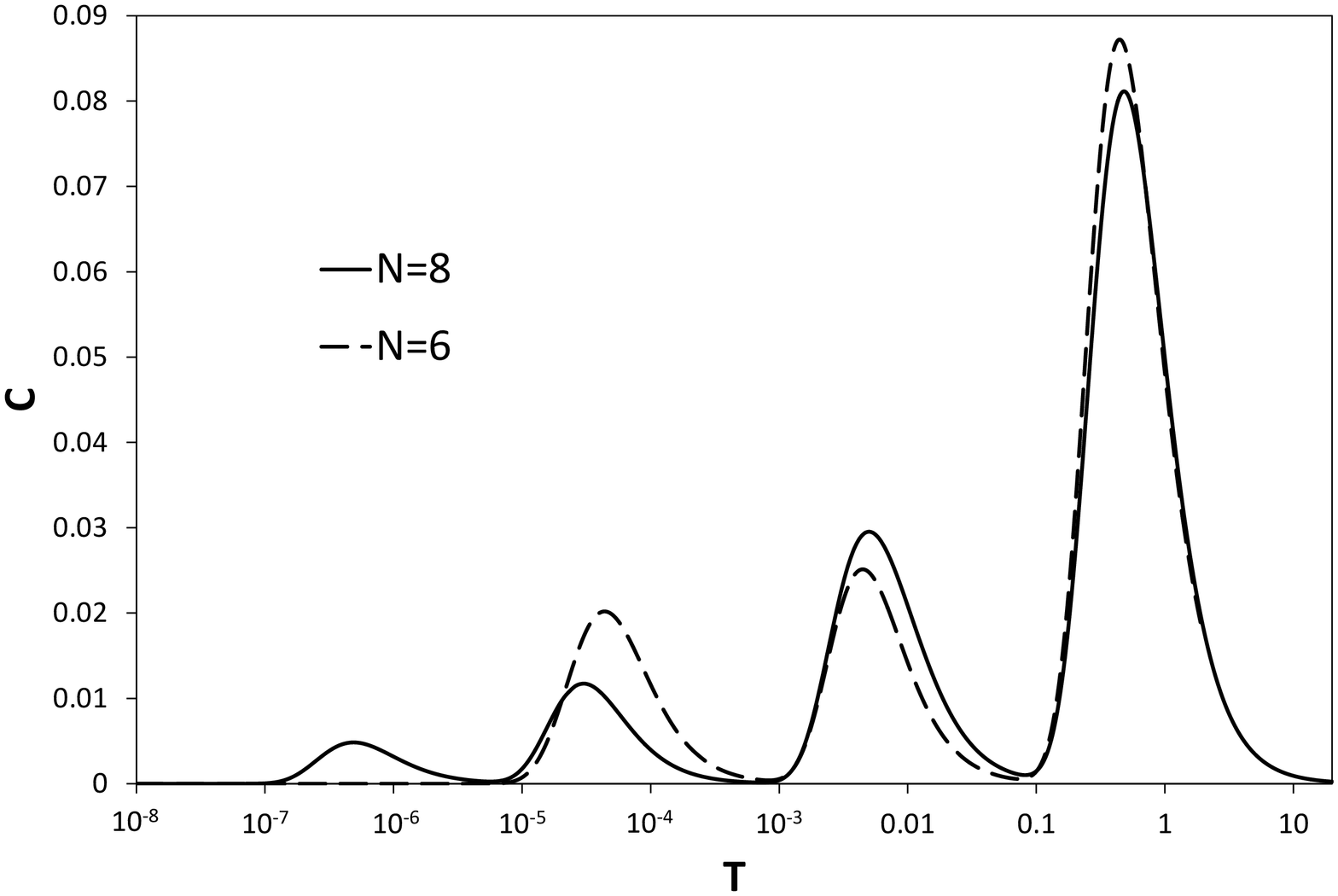}
\caption{The dependence of specific heat on the temperature of
normalized Hamiltonian (\ref{38}) for $g=0.1$ and $N=6,8$.}
\label{C_A10_N8N6}
\end{figure}

The chain consisting of $n$ triangles has $n$ peaks in the
dependence $C(T)$. Thus, the number of peaks grows linearly with
the system size $N$. It is illustrated in Fig.\ref{C_A10_N8N6},
where the specific heat $C(T)$ is shown for $N=6$ and $N=8$ with
$g=0.1$. As it can be seen from Fig.\ref{C_A10_N8N6} the number of
peaks is three (four) for $N=6 (8)$.

Since the $m$-th peak arises from the part of the spectrum with
$E\sim g^{2m}$, the maximum of the $m$-th peak takes place at
\begin{equation}
T_{m}\sim \left( cg\right) ^{2m}  \label{41}
\end{equation}%
with some constant $c$.

If the value $g$ is small the corresponding temperature $T_{m}$ is
very small as well. For this reason we could not represent in
Fig.\ref{C_A10} all feasible peaks for $N=16$ and $N=20$ because
the temperatures $T_6$ and $T_7$ for $g=0.1$ are out of the
accuracy of the ED calculations.

As it follows from the above the lowest peak should occur at
$T_{n}\sim (cg)^{2n}$. This allows us to write the finite-size
scaling parameter in the form
\begin{equation}
y=\frac{\ln T}{N\ln g}  \label{42}
\end{equation}

The dependence of the entropy per site on temperature for small
$g$ has stair-step behavior as shown in Fig.\ref{S_A10}. These
stair steps lie in between the limiting values $s=\ln 2$ for $T\to
\infty $ and $s=\frac{1}{2}\ln 2$ for $T\to 0$.

As it can be seen from Figs.\ref{C_A10} and \ref{S_A10} the data
of $C(T)$ and $s(T)$ for $N=16$ and $N=20$ deviate from each other
for $T<T_{0}=10^{-6}$ but they are indistinguishable for
$T>T_{0}$. This indicates that the obtained finite-size data
correctly describe the thermodynamic limit for temperatures
$T>T_{0}$. For example, at least three low-$T$ maxima for $C(T)$
and three stair-steps in the dependence $s(T)$ remain the shape at
$N\to \infty$ as in Figs.\ref{C_A10} and \ref{S_A10}. The levels
of the entropy $s(T)$ on these three steps testify that the number
of states in sectors is: $4^{n}$ states with $E\sim 1$; $3^{n}$
states with $E\sim g^{2} $; $2.62^{n}$ states with $E\sim g^{4}$;
$2.45^{n}$ states with $E\sim g^{6}$.

The deviation of the data for $N=16$ and $N=20$ in the region
$T<T_{0}$ means that the finite-size effects become essential for
$T<T_{0}$ and the correct description of the thermodynamics in
this temperature region needs more large systems. Nevertheless,
the multi-scale structure of the spectrum for $\Delta _{1}\gg 1$
will lead certainly to the existence of $n$ maxima in $C(T)$ and
$n$ stair-steps in $s(T)$ for chains with $n$ triangles.

\begin{figure}[tbp]
\includegraphics[width=3in,angle=-90]{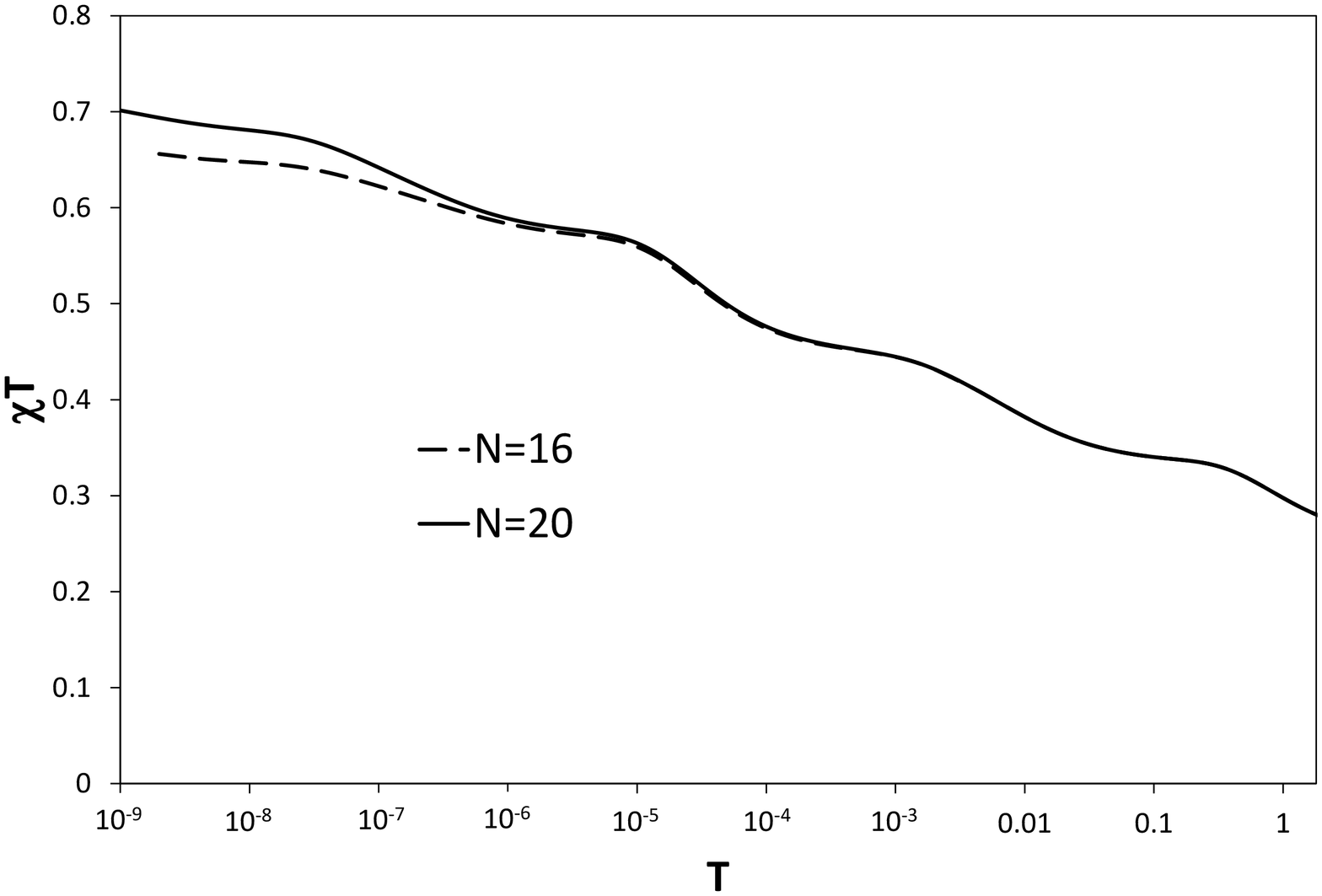}
\caption{The dependence of magnetic susceptibility on the
temperature of normalized Hamiltonian (\ref{38}) for $g=0.1$ and
$N=16,20$.} \label{chi_A10}
\end{figure}

The temperature dependence of $\chi T$ ($\chi $ is the uniform
magnetic susceptibility per spin) has a stair-like behavior
similar to the dependence $s(T)$. The numerical data show that the
quantity $\chi T$ tends to a finite value depending on $N$ at
$T\to 0$ (see Fig.\ref{chi_A10}). This finite value is related to
an average of square of $S_{total}^{z}$ over the degenerated
ground state as
\begin{equation}
\chi T=\left\langle (S^{z})^{2}\right\rangle _{0} \label{chiTSz}
\end{equation}

Evaluating this average with use of Eq.(\ref{17}) we obtain
\cite{KDNDR}

\begin{equation}
(\chi T)_{n}= \frac{N}{48},\qquad N\to \infty \label{chiTN48}
\end{equation}

According to this equation $\chi T$ is proportional to $N$ for
$N\gg 1$. It means that $\chi T$ diverges at $T\to 0$ in the
thermodynamic limit. The behavior of $\chi T$ for large $N$ and
small $T$ depends on the scale variable given by Eq.(\ref{42}). We
will determine the asymptotic behavior of $\chi T$ at $T\to 0$
using the following estimations. We assume that the
stair-step-like behavior as in Fig.\ref{chi_A10} remains in the
limit $T\to 0$, so that all steps becomes equal in this limit.
Then, for the determination of the low-temperature limit of the
susceptibility we need to find the height and width of the steps.
According to Eq.(\ref{chiTN48}) the total height is $(\chi
T)_{n}=\frac{n}{24}$. This means that each additional triangle in
the system leads to the additional stair-step of the height
\begin{equation}
h_{step}=(\chi T)_{n+1}-(\chi T)_{n}=\frac{1}{24} \label{43}
\end{equation}

The width of each stair-step can be estimated as the distance
between neighbor peaks in the specific heat, which is
\begin{equation}
w_{step}=\ln \frac{T_{m}}{T_{m+1}}=2\ln \left( cg\right)
\label{44}
\end{equation}

The low-temperature limit of the susceptibility can be calculated
as envelope of all stair-steps, so that for $\chi T$ we obtain
\begin{equation}
\chi T=\frac{h_{step}}{w_{step}}\ln T  \label{45}
\end{equation}%

Substituting Eqs.(\ref{43}) and (\ref{44}) into Eq.(\ref{45}) we
obtain the low-temperature asymptotic of the susceptibility
\begin{equation}
\chi =\frac{\ln T}{48T\ln \left( cg^{2}\right) }  \label{46}
\end{equation}

Susceptibility Eq.(\ref{46}) can be rewritten in the scaling form
which takes into account the finite-size effects:
\begin{equation}
\chi =\frac{N}{48T}f(y)  \label{47}
\end{equation}%

Here the scaling function $f(y)$ of the scaling variable
Eq.(\ref{42}) has the limits $f(y)=y$ for $y\ll 1$ and $f(\infty
)=1$. Thus, in the thermodynamic limit $\chi\sim \ln T/T$ at $T\to
0$.

\subsection{F-AF delta-chain with $\Delta_1\ll 1$}

In the limit $\Delta_1=0$ model (\ref{13}) reduces to quantum
ferromagnet on basal sites and non-interacting apical spins. For
$\Delta _{1}\ll 1$ it is convenient to normalize the Hamiltonian
(\ref{13}) in a form
\begin{equation}
\Delta_1\hat{H}=\hat{H}_{0}+\hat{V} \label{48}
\end{equation}%
where $\hat{H}_{0}$ is the Hamiltonian of the isotropic
ferromagnetic basal spin chain and $\hat{V}$ is the basal-apical
interaction
\begin{eqnarray}
\hat{H}_{0} &=&\frac{1}{2} \sum \left(
s_{2i-1}^{x}s_{2i+1}^{x}+s_{2i-1}^{y}s_{2i+1}^{y}
-s_{2i-1}^{z}s_{2i+1}^{z}+\frac{1}{4}\right)
 \nonumber \\
\hat{V} &=&-\Delta_1 \sum \left(
s_{i}^{x}s_{i+1}^{x}+s_{i}^{y}s_{i+1}^{y}\right) +\Delta_1^2
\hat{H}_{I}
\end{eqnarray}

The ground state with zero energy of $\hat{H}_{0}$ is $(n+1)$-fold
degenerated and all eigenfunctions of $\hat{H}_{0}$ do not depend
on the spin state of the apical subsystem. Therefore, the
degeneracy of the ground state of model (\ref{48}) at $\Delta_1=0$
in the spin sector $S_{total}^{z}=n-k$ is
\begin{equation}
G_{n}(k)=\sum_{m=0}^{k}C_{n}^{m}
\end{equation}%
and the total number of the ground states at $\Delta _{1}=0$ is
\begin{equation}
W=2^{n}(n+1)
\end{equation}

The perturbation $\hat{V}$ lifts the degeneracy partly and the
number of levels which split off from the ground states for fixed
$S^{z}$ is $\sum_{m=0}^{k-1}C_{n}^{m}$. These levels form the
spectrum of low-energy excitations of Hamiltonian (\ref{48}) for
$\Delta _{1}\ll 1$. In the second order in $\Delta_1$ the lowest
energy (the gap) in the spin sector $S^{z}=n-k$ for $0\leq k\leq
\frac{n}{2}$ is
\begin{equation}
E(k)=\Delta _{1}^{2}(1-\frac{2(k-1)}{n})
\end{equation}

\begin{figure}[tbp]
\includegraphics[width=3in,angle=-90]{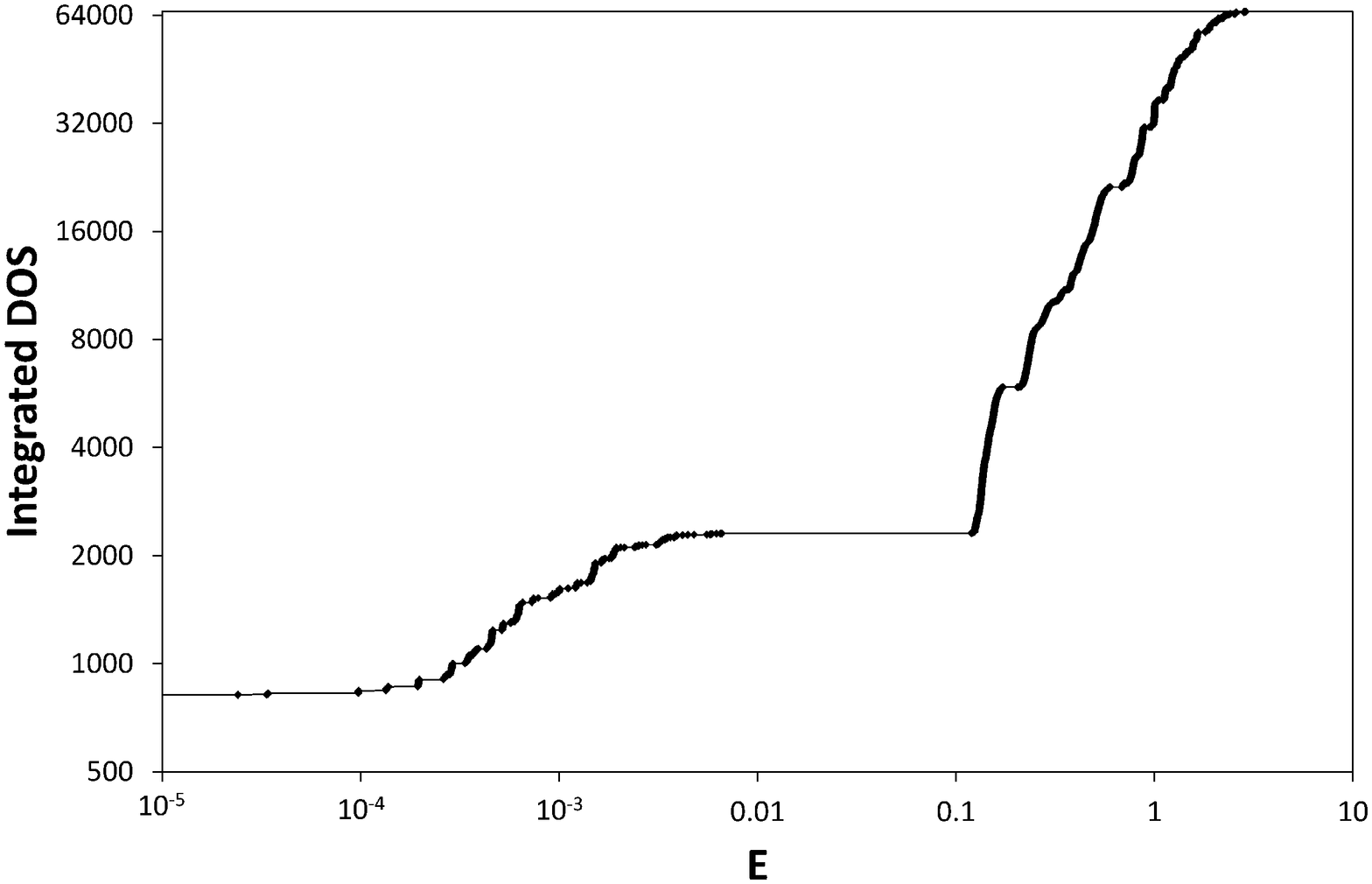}
\caption{The integrated density of states of normalized
Hamiltonian (\ref{48}) for $\Delta_1=0.05$ and $N=16$.}
\label{DOS_A01}
\end{figure}

\begin{figure}[tbp]
\includegraphics[width=3in,angle= -90]{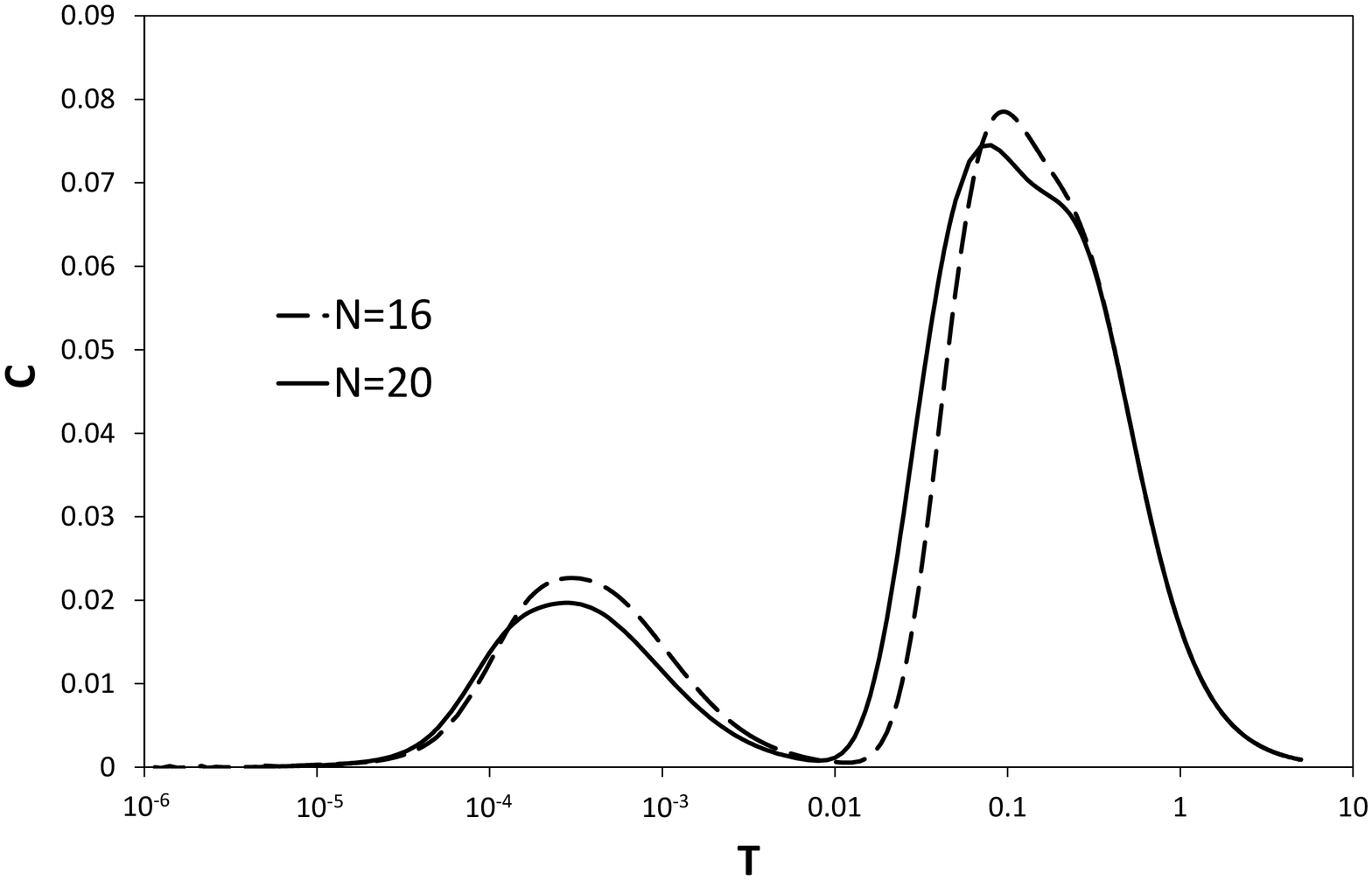}
\caption{The dependence of specific heat on the temperature of
normalized Hamiltonian (\ref{48}) for $\Delta _{1}=0.05$ and
$N=16,20$.} \label{C_A01}
\end{figure}

For example, the gap for the one-magnon excitations is
$E(1)=\Delta_1^2$. This agrees with the energy of the upper branch
of the one-magnon states Eq.(\ref{2}) at $q=\pi$. As to the gaps
in the spin sectors $\frac{n}{2} <k\leq n$, the numerical
calculations of finite chains show that they are also proportional
to $\Delta_1^2$. Therefore, the spectrum of model (\ref{48}) at
$\Delta _{1}\ll 1$ has two-scale structure with the energies
$E\sim\Delta_1^2$ and $E\simeq 1$. It is illustrated in
Fig.\ref{DOS_A01} where the integrated density of states is shown
for $N=16$ and $\Delta_1=0.05$. The two-scale spectrum for $\Delta
_{1}\ll 1$ is in contrast with the multi-scale spectrum for
$\Delta _{1}\geq 1$. Such two-scale structure of the spectrum
leads to the temperature dependence of the specific heat with two
peaks as shown in Fig.\ref{C_A01} for $\Delta _{1}=0.05$. This
behavior is in contrast with multi-peaks form of $C(T)$ for
$\Delta _{1}\gg 1$.

\begin{figure}[tbp]
\includegraphics[width=3in,angle=-90]{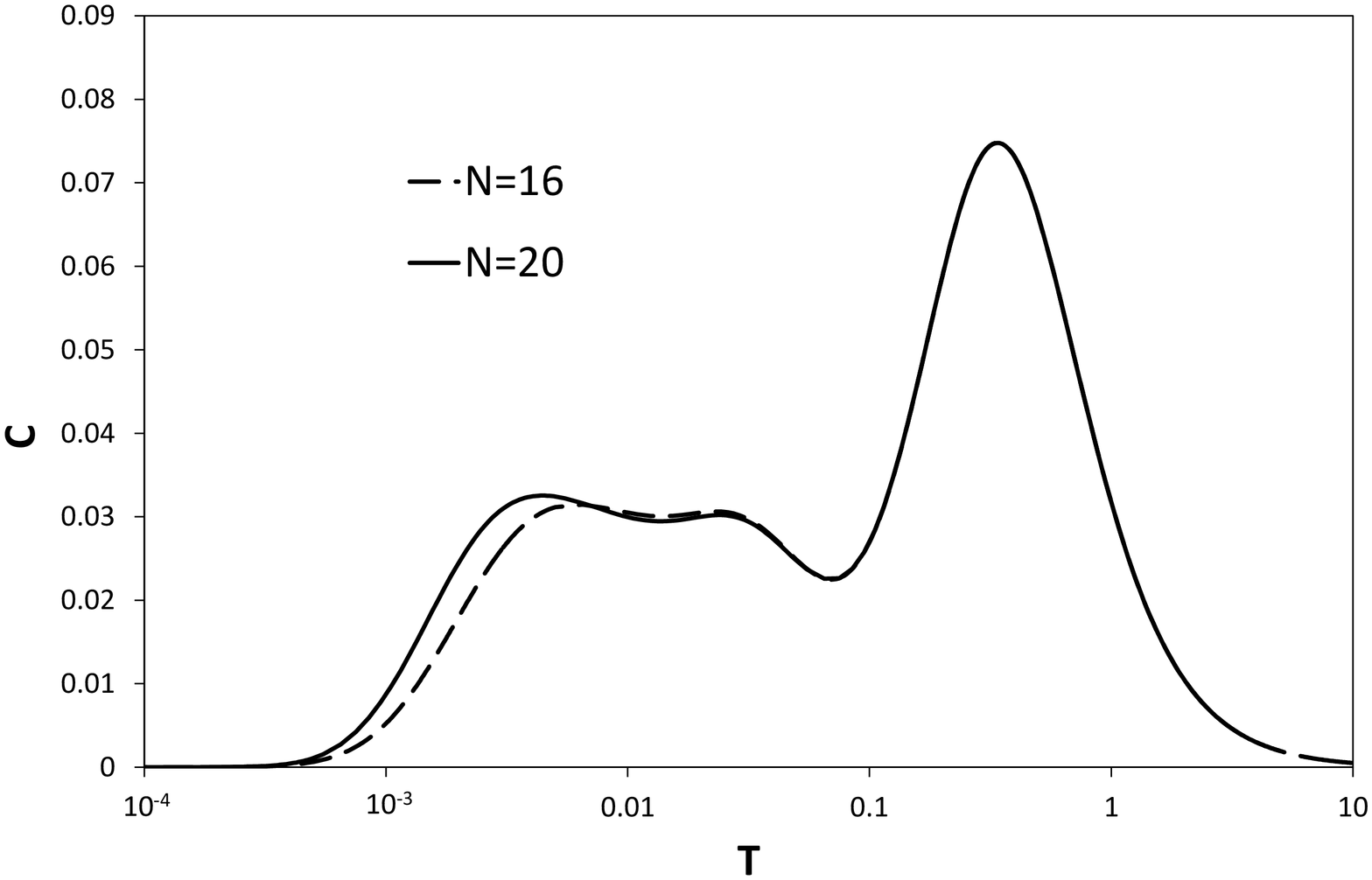}
\caption{The dependence of specific heat on the temperature of
normalized Hamiltonian (\ref{48}) for $\Delta_1=0.5 $ and
$N=16,20$.} \label{C_A1}
\end{figure}

\subsection{F-AF delta-chain with intermediate value of
$\Delta_1$.}

As follows from the results obtained above, the structure of the
spectrum of model (\ref{13}) as well as the low-temperature
thermodynamics is qualitatively different in the limiting cases
$\Delta _{1}\gg 1$ and $\Delta _{1}\ll 1$. When the parameter
$\Delta _{1}$ is neither large nor small the situation is less
clear.

\begin{figure}[tbp]
\includegraphics[width=3in,angle=-90]{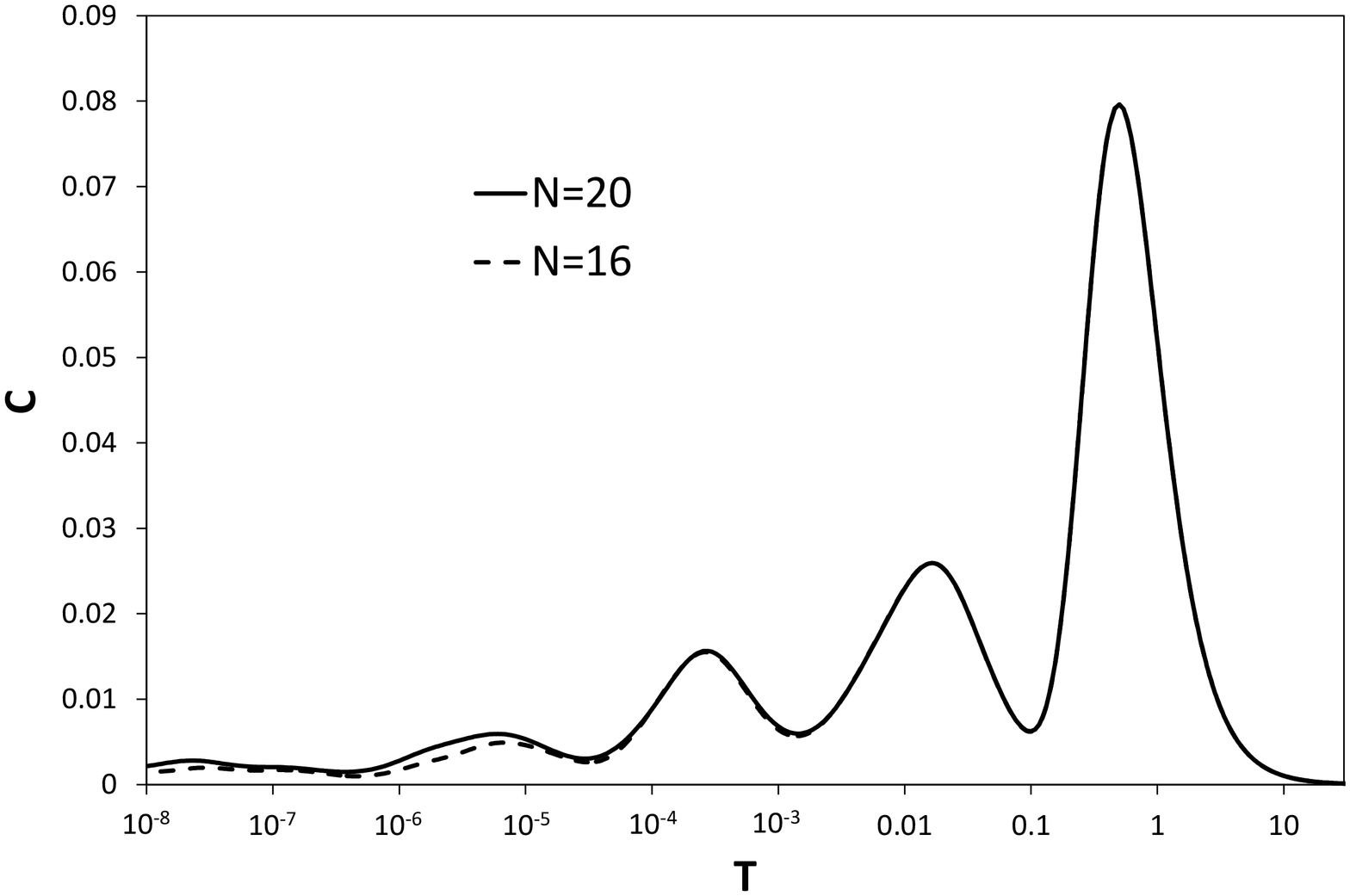}
\caption{The dependence of specific heat on the temperature of
normalized Hamiltonian (\ref{38}) for $g=0.2$ ($\Delta_1=2.5$) and
$N=16,20$.} \label{C_A5}
\end{figure}

\begin{figure}[tbp]
\includegraphics[width=3in,angle=-90]{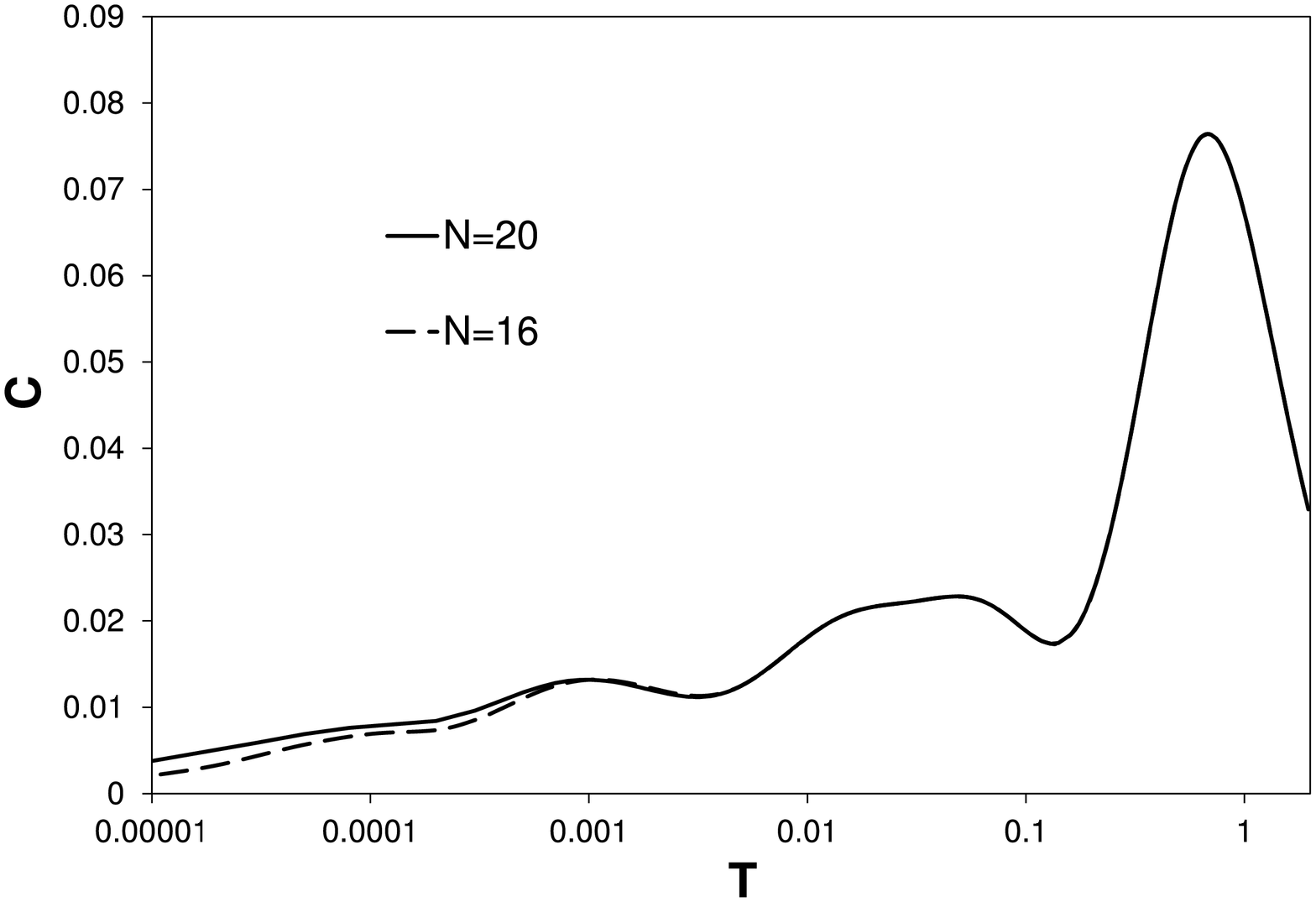}
\caption{The dependence of specific heat on the temperature for
$\Delta_1=1$ and $N=16,20$.} \label{C_A2}
\end{figure}

Let the parameter $\Delta _{1}$ moves down from the limiting case
$\Delta _{1}\gg 1$. Then the ordered partition of the spectrum is
smeared and the multi-scale structure of spectrum becomes less
obvious. Nevertheless for $\Delta _{1}\geq 1$ the low-temperature
behavior of the thermodynamic quantities remains qualitatively
similar to that for $\Delta _{1}\gg 1$. For example, the
temperature dependence of $C(T)$ for the F-AF delta-chain with
$\Delta_1=2.5$ and $\Delta _{1}=1$ is shown in Fig.\ref{C_A5} and
Fig.\ref{C_A2}. The specific heat is characterized by the
existence of the low-T maxima, though they are not so distinctive
as in the case $\Delta _{1}=5$ (Fig.\ref{C_A10}). The reason of
this fact is that the finite-size effects become more pronounced
when $\Delta_1$ decreases. For example, the temperature $T_{0}$
below which the data for $N=16$ and $N=20$ start to deviate from
each other is $T_{0}\sim 10^{-6}$, $10^{-4}$, $10^{-3}$ for
$\Delta _{1}=5$, $\Delta _{1}=2.5$, $\Delta _{1}=1$,
correspondingly. Nevertheless, we expect that the number of low-T
peaks of $C(T)$ is proportional to $N$. The behavior of other
thermodynamic quantities such as the entropy $s(T)$ and $\chi
T(T)$ for $\Delta _{1}=2.5$ and $\Delta _{1}=1$ is qualitatively
similar to those for $\Delta _{1}\gg 1$. This suggests that model
(\ref{13}) with $\Delta _{1}\gg 1$ is the generic one for all
values of the anisotropy parameter in the range $\Delta _{1}\geq
1$.

However, for $0<\Delta_1<1$ the low-T behavior of the
thermodynamics changes qualitatively. As it can be seen from
Fig.\ref{C_A01} and Fig.\ref{C_A1} the specific heat $C(T)$ has
more complicated form. In addition to the broad maximum at
$T\simeq 1$ there is only one (or two) low-T maximum.
Unfortunately, the finite-size effects are rather large for
$0<\Delta _{1}<1$ and the temperature $T_{0}$ is $T_{0}\simeq
0.01$ and $T_{0}\simeq 0.1$ for $\Delta _{1}=0.5$ and $\Delta
_{1}=0.05$. This complicates considerably the determination of the
behavior in the thermodynamic limit. All the same, we believe that
multi-peak structure for $C(T)$ or many stair-steps one for $s(T)$
does not appear for $0<\Delta _{1}<1$ and the model with $\Delta
_{1}\ll 1$ is a generic one in the range $0<\Delta _{1}<1$.

Thus, we suggest that the value $\Delta _{1}=1$ divides the
regions of the parameter $\Delta _{1}$ with the qualitatively
different behavior of the low-temperature thermodynamics. The
possible argument in favor of this assumption is the fact that
this model describes the transition line between the ground state
phases of different types for $\Delta _{1}\geq 1$ and $\Delta
_{1}<1$ (see Fig.\ref{phases}).

\section{Summary}

We have studied the ground state and the low-temperature
thermodynamics of the delta-chain with anisotropic F and AF
interactions. At definite relations between values of these
interactions the lowest branch of the one-magnon states is
dispersionless which means the existence of exact localized
eigenfunctions. If the energy of the localized states is zero
(lowest energy) then they are the ground states. In this case the
model depends on a single parameter which can be chosen as the
anisotropy of basal-apical spin interaction $\Delta_1$. When
$\Delta_1=\infty$ the model reduces to the Ising model with equal
F and AF interactions while for $\Delta_1=0$ it is the isotropic
ferromagnet on the basal chain and independent spins on the apical
sites. Remarkably, the ground state degeneracy is the same for all
$0<\Delta_1<\infty$ (excluding some special values of $\Delta_1$).
The degeneracy is macroscopic and leads to finite residual entropy
at $T=0$. In the limiting cases $\Delta_1=0$ and $\Delta_1=\infty
$ the ground state degeneracy is even larger and for finite
$\Delta_1$ it is partially lifted.

For $\Delta_1 \gg 1$ the excitation spectrum has multi-scale
structure and is rank-ordered in powers of small parameter
$\Delta_1^{-2}$. The number of sections of the spectrum is equal
to the number of triangles of the delta-chain and the energy of
the levels in the $m$-th section $E\sim\Delta_1^{-2m+2}$
($m=1\ldots N/2$). The origin of such exponentially low energy
levels is the fact that the $m$-magnon bound complex in this
system has the energy $E_m\sim\Delta_1^{-2m+2}$. Each $m$-th
section of the spectrum is responsible for the appearance of
$m$-th peak in the dependence $C(T)$. Thus, the number of the
peaks in the specific heat grows with the length of the chain.
Similarly, such multi-scale structure of the spectrum determines
the characteristic features of the dependence of $s(T)$ and
$\chi(T)$. Numerical calculations by the ED of finite delta-chains
for smaller values of $\Delta_1$ (up to the isotropic point
$\Delta_1=1)$ show that the behavior of the thermodynamic
quantities is qualitatively similar to that for $\Delta_1 \gg 1$.

For $\Delta_1 \ll 1$ the spectrum has two-scale structure in
contrast with the case $\Delta_1 \gg 1$. It leads to only one
low-temperature maximum in $C(T)$. According to the numerical
calculations such feature in the behavior of $C(T) $ survives in
the region $\Delta_1<1$. Thus, the isotropic point $\Delta_1=1$
separates the region of $\Delta_1$ on two parts with the
qualitatively different behavior of low-temperature
thermodynamics. We note that this conclusion is based on numerical
calculations and needs more rigorous analysis.

The main points of our study can be extended to delta-chains with
arbitrary spin values because the condition on exchange
interactions for existence of localized magnons does not depend on
the value of spin $s$.

\begin{acknowledgments}
We would like to thank J.Richter for valuable comments on the
manuscript. The numerical calculations were carried out with use
of the ALPS libraries \cite{alps}.
\end{acknowledgments}


\begin{thebibliography}{99}

\bibitem{flat2008} O.\ Derzhko, J.\ Richter, M.\ Maksymenko,
Int. J. Modern Phys. \textbf{29}, 1530007 (2015).

\bibitem{Wu} C.\ Wu, D.\ Bergman, L.\ Balents, and S.\ Das Sarma,
Phys.Rev.Lett. \textbf{99}, 070401 (2007).

\bibitem{Wang} Y.-F.\ Wang, Z.-C.\ Gu, C.-D.\ Gong and D.N.\ Sheng, Phys.Rev.Lett.
\textbf{107, }146803 (2011).

\bibitem{Sun} K.\ Sun, Z.\ Gu, H.\ Katsura and S.\ Das Sarma, Phys.Rev.Lett. \textbf{%
106, }236803 (2011).

\bibitem{Shulen} J.\ Richter, O.\ Derzhko and J.\ Schulenburg, Phys.Rev.Lett.
\textbf{93, }107206 (2004).

\bibitem{Mak} M.\ Maksymenko, A.\ Honecker, R.\ Moessner, J.\ Richter, and
O.\ Derzhko, Phys.\ Rev.\ Lett.\ \textbf{109}, 096404 (2012).

\bibitem{Zhit} M.\ E.\ Zhitomirsky and H.\ Tsunetsugu, Phys.Rev.B 70, 100403
(2004).

\bibitem{Schmidt} J.\ Schnack, H.-J.\ Schmidt, J.\ Richter and J.\
Schulenberg, Eur.\ Phys.\ J.\ B \textbf{24}, 475 (2001).

\bibitem{Honecker} J.\ Richter, J.\ Schulenburg, A.\ Honecker, J.\ Schnack,
and H.J.\ Schmidt, J.\ Phys.: Condens.\ Matter \textbf{16}, S779 (2004).

\bibitem{Derzhko2004} O.\ Derzhko and J.\ Richter, Phys.\ Rev.\ B \textbf{70}%
, 104415 (2004).

\bibitem{Zhitomir} M.E.\ Zhitomirsky and H.\ Tsunetsugu, Progr. Theor. Phys.
Suppl. \textbf{160}, 361 (2005).

\bibitem{Sen} D.\ Sen, B.S.\ Shastry, R.E.\ Walsteadt and R.\ Cava, Phys.\
Rev.\ B \textbf{53} ,6401 (1996).

\bibitem{Nakamura} T.\ Nakamura and K.\ Kubo, Phys.\ Rev.\ B \textbf{53},
6393 (1996).

\bibitem{Blundell} S.A.\ Blundell and M.D.\ Nuner-Reguerio, Eur.\ Phys.\ J.\
B \textbf{31}, 453 (2003).

\bibitem{Inagaki} Y.\ Inagaki, Y.\ Narumi, K.\ Kindo, H.\ Kikuchi, T.\
Kamikawa, T.\ Kunimoto, S.\ Okubo, H.\ Ohta, T.\ Saito, H.\ Ohta, T.\ Saito,
M.\ Azuma, H.\ Nojiri,, M.\ Kaburagi and T.\ Tonegawa, J.\ Phys.\ Soc.\
Jpn.\ \textbf{74}, 2831 (2005).

\bibitem{Kaburagi} M.\ Kaburagi, T.\ Tonegawa and M.\ Kang, J.Appl.Phys.
\textbf{97}, 10B306 (2005).

\bibitem{ruiz} C.\ Ruiz-Pérez, M.\ Hernández-Molina, P.\ Lorenzo-Luis,
F.\ Lloret, J.\ Cano, and M.\ Julve, Inorg. Chem. \textbf{39} 3845
(2000).

\bibitem{Tonegawa} T.\ Tonegawa and M.\ Kaburagi, J.\ Magn.\ Magn.\
Materials, \textbf{272}, 898 (2004).

\bibitem{KDNDR} V.\ Ya.\ Krivnov, D.\ V.\ Dmitriev, S.\ Nishimoto
S.-L.\ Drechsler, and J.\ Richter, Phys. Rev. B \textbf{90},
014441 (2014).

\bibitem{Derzhko2008} J.\ Richter, O.\ Derzhko and A.\ Honecker,
J.Modern Phys. B\textbf{22}, 4418 (2008).

\bibitem{hub1} O.\ Derzhko, A.\ Honecker, and J.\ Richter,
Phys. Rev. B \textbf{76}, 220402 (2007).

\bibitem{hub2} M.\ Maksymenko, A.\ Honecker, R.\ Moessner, J.\ Richter,
and O.Derzhko, Phys. Rev. Lett. \textbf{109}, 096404 (2012).

\bibitem{alps}  F.\ Alet et al., J.Phys.Soc.Jpn.Suppl. \textbf{74}, 30 (2005).

\end{thebibliography}
\end{document}